\documentclass[%
 12pt,
 tightenlines,
 superscriptaddress,
 nofootinbib,
 notitlepage,
 bibnotes,
 amsmath,amssymb,
 aps,
 prd,
]{revtex4-1}
\usepackage{units}
\usepackage{graphicx}
\usepackage{dcolumn}
\usepackage{bm}
\usepackage{leftidx}
\usepackage{amsfonts,amsmath,epsfig,bm,physics}
\usepackage[dvipsnames]{xcolor}
\usepackage{graphicx,hyperref}
\usepackage{booktabs}
\newcommand\numberthis{\addtocounter{equation}{1}\tag{\theequation}}
\usepackage{esvect}
\usepackage[caption=false]{subfig}
\hypersetup{
    colorlinks,
    citecolor=red,
    filecolor=red,
    linkcolor=Maroon,
    urlcolor=Maroon
}
\usepackage{mathtools}          

\begin{document}

\preprint{APS/123-QED}

\title{Impact of nonlinear overdensity statistics on primordial black hole abundance}

\author{Rafid Mahbub}
\email{mahbu004@umn.edu}
\affiliation{%
 School of Physics \& Astronomy, University of Minnesota,\\
 Minneapolis, MN 55455, USA 
}%

\date{\today}

\begin{abstract}
It has been recently established that, if the nonlinear relationship between the overdensity perturbations and the curvature perturbations are taken into account, non-Gaussianity is introduced in the overdensity statistics which alters the expected primordial black hole abundance. This is explored by using the nonlinear relationship between the overdensities and curvature perturbations up to second order where a negative skewness and positive kurtosis aims at lowering and increasing the abundance while an abundance comparable to Gaussian perturbations is obtained by adjusting the amplitude of the curvature power spectrum. The effects of the nonvanishing skewness and kurtosis are studied using a toy model Dirac delta and lognormal curvature power spectra as well as one obtained from an $\alpha-$attractor model capable of primordial black hole production. Finally, the nonlinear calculations using Press-Schechter are compared with peaks theory.
\end{abstract}

\maketitle

\tableofcontents

\section{Introduction}

The discovery in 2015 of the merger of two $\sim 30 M_{\odot}$ black holes brought about the era of gravitational wave (GW) astronomy along with renewed interest in primordial black holes (PBHs) \cite{abbott}. PBHs are hypothetical black holes that were theoretically conceived of in the 70s by Hawking who, later with Carr, produced some of the earliest works on this subject \cite{hawking,hawking2,carr1}. While PBHs are indeed black holes, they do not form from stellar collapse. They are thought to have formed in the very early radiation dominated (RD) era of the Universe from the collapse of overdense regions. These overdense regions could have been formed through various mechanisms (e.g. through a softened equation of state due to some first order phase transition \cite{Jedamzik:1999am}), though the prevailing and most widely accepted one is that these overdensities were seeded by inflation. However, to bring about collapse in the radiation plasma of the early Universe to form black holes, these overdensities necessarily need to be large and rare and, hence, cannot be explained by slow-roll models. Such inflation models predict a nearly scale invariant curvature power spectrum of $\mathcal{P}_{\zeta}\sim 10^{-9}$ at cosmologically observable scales $k_{\text{CMB}}=\unit[0.05]{Mpc^{-1}}$ which decreases as a power of $k^{n_{s}-1}$ with scale, where $n_{s}\approx 0.96$ is the scalar spectral index \cite{Akrami:2018odb}. The nature of the curvature power spectrum is largely unconstrained at very small scales ($k\gg k_{\text{CMB}}$) and inflationary scenarios can be constructed where the curvature power spectrum experiences a large amplification at some small scale. Such scenarios have been studied in great detail recently under the context of ultra slow-roll (USR) inflation models \cite{ballesteros,dalianis,Gong:2017qlj}. Characteristic of such models is an inflection point in the inflaton potential where the inflaton slows down rapidly, giving rise to the required amplification of $\mathcal{P}_{\zeta}$. Although there has been no evidence of PBHs as of yet, they are nevertheless important in cosmology. They possess the properties of cold dark matter candidates of being nonrelativistic and collisionless. Moreoever, since they should have formed before the Big Bang Nucleosynthesis, PBHs avoid the constraints imposed on baryonic dark matter. Formation of PBHs can also be used to explain the existence of super massive black holes. Nevertheless, there exists numerous constraints because of which the scenarios where PBHs comprise the totality of cold dark matter (CDM) have been ruled out in most of the viable mass ranges \cite{Carr:2016drx,Carr:2020gox,sasaki2}.  \\\\
\indent The curvature perturbations that are generated during inflation become frozen when they become superhorizon. These eventually reenter the horizon and produce the density fluctuations in the matter-energy content of the Universe after inflation. It has been standard practice in literature to use a linear relationship between the curvature perturbations $\zeta(\bm{x},t)$ and the overdensities $\delta(\bm{x},t)$ given by \cite{Young:2014ana,Green:2004wb}
\begin{equation}\label{eq:delta_lin}
\delta = \frac{4}{9}\left( \frac{1}{aH} \right)^{2}\nabla^{2}\zeta
\end{equation}

Consequently, this made Fourier analysis an easy task since calculating the curvature power spectrum by solving the Mukhanov-Sasaki equation can be accomplished in a relatively straightforward manner. However, the relationship between these two quantities is highly nonlinear and one cannot neglect higher order effects in the study of PBHs.\footnote{This makes Fourier analysis a more involved process as a direct and simple relationship between $\zeta(k)$ and $\delta(k)$ cannot be established.} As such, it has been shown that, if this nonlinear relationship is taken into account, the overdensities unavoidably develop a non-Gaussianity in their distribution even in the absence of primordial non-Gaussianities in $\zeta$. Furthermore, it has also been demonstrated that such non-Gaussianities hinder the PBH formation process and would require perturbations with higher amplitudes to yield abundances comparable to the Gaussian case \cite{DeLuca:2019qsy,Kawasaki:2019mbl,Young:2019yug}. Specifically, in \cite{Kawasaki:2019mbl} it was shown that upto $\mathcal{O}(\zeta^{2})$, the overdensities develop a negative skewness which lowers the PBH abundance. However, a similar manner of calculation reveals that the kurtosis of the distribution of the overdensity fields is positive. As a result, the area under the tail of the underlying probability distribution (PDF) is greater than the Gaussian case, implying larger PBH abundance. \\\\
\indent In this work, the effects of the nonlinear relationship between $\zeta(\bm{x},t)$ and $\delta(\bm{x},t)$ are studied along with how a nonvanishing skewness and kurtosis of the underlying PDF of the PBH formation fraction modifies the abundance. The calculations use the concept of volume-averaged overdensities $\bar{\delta}$ expanded up to second order in curvature perturbations. This has the benefit that $\bar{\delta}$ can be directly related to the compaction function, which establishes collapse criteria more accurately. In Sec. (\ref{sec:PBH_formation}), the PBH formation criteria will be briefly reviewed using well-established results from gradient expansion formalism and compaction function. In Sec. (\ref{sec:overdensity_statistics}), using the volume-averaged overdensities, the first three nonvanishing cumulants of the overdensities will be considered which, using Sec. (\ref{sec:PS_mod}), will then be used in Sec. (\ref{sec:PBH_abundance}) to study the impact of nonlinearities on PBH abundance using toy model $\mathcal{P}_{\zeta}$ like Dirac delta and lognormal and one obtain from a particular realization of $\alpha-$attractor inflation. Finally, in Sec. (\ref{sec:peaks_compare}), the results from Press-Schechter will be compared to an optimized calculation based on peaks theory. In this work, $c=M_{\text{p}}=(8\pi G)^{-1}=1$ unless otherwise specified.

\section{PBH formation criteria}
\label{sec:PBH_formation}
PBHs are expected to form when certain overdense regions exceed some threshold overdensity $\delta_{\text{th}}$. When such a situation occurs, these regions are able to overcome outward pressure forces from the radiation plasma and gravitationally collapse. One of the earliest estimates of this threshold was derived by Carr using simple, nonrelativistic Jeans instability arguments in an expanding Universe where it was shown that $\delta_{\text{th}}\approx1/3$ \cite{Yokoyama:1995ex}. However, this was rather simplistic and later numerical relativity simulations have produced different sets of values which could act as more accurate determinants of threshold overdensities which, nevertheless, were dependent on the simulation details. We assume that the Universe is spherically symmetry and described by the following metric 
\begin{equation}
ds^{2}=-dt^{2}+a(t)^{2}e^{2\zeta(r)}\left( dr^2 +r^2 d\Omega^2 \right)
\end{equation} 
where $a(t)$ is the scale factor and $\zeta(r)$ is the curvature perturbation.\footnote{In linear cosmological perturbation theory, the term $e^{2\zeta}$ would have been expanded into $1 + 2\zeta +\mathcal{O}(\zeta^2)$.} As mentioned earlier, PBHs form from the collapse of highly overdense regions which were initially much larger than the size of the comoving Hubble horizon $(aH)^{-1}$ where the separate universe assumption can be applied \cite{Rigopoulos:2003ak}. There, one can make use of the gradient expansion approach \cite{Lyth:2004gb,Harada:2015yda} where a fictitious parameter $\epsilon=k/aH$ is introduced in front of spatial derivatives in the Einstein equations such that $\partial_{i}\rightarrow \epsilon\partial_{i}$. On super Hubble scales, $\epsilon \ll 1$ and relevant quantities can be expanded in powers of $\epsilon$ upto the desired order and then $\epsilon$ can be set to unity (corresponding to horizon crossing). To the leading order in $\epsilon$ and in the comoving gauge, the overdensities then can be expressed as 
\begin{equation}
\frac{\delta\rho}{\bar{\rho}}\equiv\delta(r,t)\simeq-\frac{8}{9}\left( \frac{1}{aH} \right)^2 e^{-5\zeta(r)/2}\nabla^{2}e^{\zeta(r)/2}
\end{equation}
where $\bar{\rho}(t)=3M_{\text{p}}^{2}H(t)^{2}$ is the energy-density of the radiation background and the only time dependence is assumed to come from the factor $a(t)H(t)$. Equation \eqref{eq:delta_lin} can be recovered by taking the linear approximation but it is easy to see that the exponential terms start producing nontrivial contributions once $\zeta\sim\mathcal{O}(1)$ which are precisely the conditions under which PBH formation takes place. It can be said that PBHs form when certain overdense regions denser than a certain threshold value collapse. A more refined criterion was introduced by Shibata and Sasaki \cite{Shibata:1999zs} through the construction of what is known as the compaction function.\footnote{Shibata and Sasaki's original formulation of the compaction function did not have the factor of two.} This is defined as twice the gravitational potential of the mass excess within a region of areal radius $R(r,t)=a(t)re^{\zeta(r)}$.\footnote{The concept of the areal radius here comes from the fact that the most general, spherically symmetric metric can be written as $ds^{2}=-A(r,t)^{2}dt^2 +B(r,t)^{2}dr^{2}+R(r,t)^{2}d\Omega^{2}$} Mathematically, it takes the following form
\begin{equation}
\mathcal{C}(r,t)\equiv \frac{2G\delta M(r,t)}{R(r,t)}=2G\frac{M_{\text{MS}}(r,t)-\bar{M}(r,t)}{R(r,t)}
\end{equation}
The mass excess appearing in the numerator is actually the difference between the Misner-Sharp mass $M_{\text{MS}}(r,t)$ and the background mass $\bar{M}(r,t)=4\pi\bar{\rho}(t)R^{3}(r,t)/3$ within the areal radius. The Misner-Sharp mass is a quasilocal mass defined for spherically symmetric spacetimes. It takes the following form \cite{Misner:1964je,Harada:2013epa}
\begin{equation}
M_{\text{MS}}=\frac{1}{2G}R(1-g^{\mu\nu}\nabla_{\mu}R\nabla_{\nu}R)
\end{equation}
where $R$ is the previously defined areal radius and should not be confused with the Ricci scalar. The Misner-Sharp mass has become useful in numerical relativity because of its utility in locating apparent horizons, required in simulations. A short calculation will show that the mass excess can be expressed as 

\begin{align*}
\delta M(r,t)&=4\pi\int_{0}^{r}dr'\frac{dR(r',t)}{dr'}R(r',t)^{2}\frac{\rho(r',t)-\bar{\rho}(t)}{\bar{\rho}(t)} \\
&=4\pi\int_{0}^{r}dr'\frac{dR(r',t)}{dr'}R(r',t)^{2}\frac{\delta\rho(r',t)}{\bar{\rho}(t)} \numberthis
\end{align*}
Now, there exists a simple expression for the compaction function in the superhorizon limit 
\begin{equation}\label{eq:compaction}
\mathcal{C}(r)=-\frac{2}{3}r\zeta'(r)\left( 2+r\zeta'(r) \right)
\end{equation}
It can be shown that the compaction function is conserved in the superhorizon limit \cite{Kehagias:2019eil,Young:2019yug}. Collapse depends on the maximization of the compaction function occuring at a certain comoving length scale defined to be $r_{m}$, such that for $\mathcal{C}'(r_{m})=0$
\begin{equation}
\zeta'(r_{m})+r_{m}\zeta''(r_{m})=0
\end{equation}
The scale $r_{m}$ can therefore be treated as a typical lengthscale that characterizes PBH forming overdensities and collapse will occur if, for some perturbation profile, $\mathcal{C}(r_{m})>\mathcal{C}_{\text{th}}(r_{m})$, where the threshold $\mathcal{C}_{\text{th}}$ is determined through numerical simulations. As the collapse takes place when the perturbations, originally superhorizon, becomes of the order of the Hubble horizon, the horizon crossing condition also needs to be specified. Working in real space, it can be expressed as $R(r_{m},t_{H})H(r_{m},t_{H})=1$ or, expressing the areal radius in terms of the local scale factor, $a(t_{H})r_{m}e^{\zeta(r_{m})}H(r_{m},t_{H})=1$. Furthermore, instead of dealing with the overdensity $\delta(r,t)$ itself, it has been suggested that a volume-averaged overdensity is more suitable. It has the added benefit that it can be related to the compaction function. Then, at the scale which maximizes the compaction function, the volume-averaged overdensity is defined by the following (evaluated at horizon crossing and suppressing the time label) \cite{Musco:2018rwt}
\begin{align*}\label{eq:volume_averaged_density}
\bar{\delta}(r_{m})&=\frac{1}{\frac{4\pi}{3}r_{m}^{3}}\int_{0}^{r_{m}}dr\; 4\pi r^{2}\frac{\delta\rho}{\bar{\rho}}\\ 
&=-\frac{2}{3}\left( 2r_{m}\zeta'(r_{m}) +r_{m}^{2}\zeta'(r_{m})^2 \right)\\ 
&=\bar{\delta}^{(1)}(r_{m})+\bar{\delta}^{(2)}(r_{m}) \numberthis
\end{align*}
where the superscripts represent order one and order two contributions to the volume-averaged overdensity. Given a perturbation profile, it can be demonstrated that $\bar{\delta}$ does not grow linearly with $\zeta$ but rather it is suppressed \cite{Kawasaki:2019mbl}. Equation \eqref{eq:volume_averaged_density} shows that, at the scale of the perturbation, $\mathcal{C}(r_{m})=\bar{\delta}(r_{m})$. In \cite{Musco:2018rwt}, $\mathcal{C}_{\text{th}}(r_{m})=\bar{\delta}_{\text{th}}(r_{m})\simeq 0.5$ was identified as a suitable threshold for PBH formation and values close to this will be used in computations in Sec. (\ref{sec:PBH_abundance}).\footnote{In fact, in \cite{Musco:2018rwt}, two thresholds were actually discussed- $0.5\leq \bar{\delta}_{\text{th}}(r_{m})\leq 2/3$ for type I PBHs and $\bar{\delta}_{\text{th}}(r_{m})>2/3$ for type II PBHs.}
\section{Statistics of the overdensity fields}
\label{sec:overdensity_statistics}
Inflation produces nearly Gaussian perturbations and the density fluctuations have been described as conforming to such a distribution which can be characterized simply by the variance of the fields. Of course, if primordial non-Gaussianities are considered one needs to also evaluate higher order cumulants to better specify the statistics of the density fields. As mentioned earlier, recently a few papers have shown that nonlinearities introduce non-Gaussianities in the overdensities even for Gaussian inflationary perturbations. Following \cite{Kawasaki:2019mbl,Liddle:2000cg}, the curvature perturbations are expanded using spherical harmonics
\begin{align}\label{eq:spherical_fourier}
\zeta(\bm{x})&=\int_{0}^{\infty}dk\sum_{l,m}\sqrt{\frac{2}{\pi}}kj_{l}(kr)Y_{m}^{l}(\hat{\bm{x}})\zeta_{lm}(k) \\ \nonumber
\langle \zeta_{lm}(k)\zeta^{*}_{l'm'}(k') \rangle&=\delta_{ll'}\delta_{mm'}\delta(k-k')\frac{2\pi^2}{k^3}\mathcal{P}_{\zeta}(k)
\end{align}
where $j_{l}(kr)$ are the spherical Bessel functions of the first kind and $Y^{l}_{m}(\hat{\bm{x}})$ are the spherical harmonics. Spherical symmetry implies that only the $l,m=0$ components survive and the sum can be removed. For the sake of brevity, $\zeta_{00}$ shall be referred to as simply $\zeta$. The variance and skewness of $\bar{\delta}$ can be computed using their regular definitions. Details on the derivation can be found in Appendix (\ref{appendix:A}). The variance, skewness and kurtosis are defined as
\begin{align*}
\sigma^{2}(r_{m})&=\langle \bar{\delta}(r_{m})^{2} \rangle -\langle \bar{\delta}(r_{m}) \rangle^{2}\numberthis \\
\tilde{\kappa}_{3}(r_{m})&=\frac{1}{\sigma(r_{m})^{3}}\left[ \langle \bar{\delta}(r_{m})^{3}\rangle-3\langle \bar{\delta}(r_{m})^{2} \rangle \langle \bar{\delta}(r_{m}) \rangle +2\langle \bar{\delta}(r_{m}) \rangle^{3} \right]\numberthis\\
\tilde{\kappa}_{4}(r_{m})&=\frac{1}{\sigma(r_{m})^{4}}\biggl[ \langle \bar{\delta}(r_{m})^{4} \rangle-4\langle \bar{\delta}(r_{m}) \rangle\langle \bar{\delta}(r_{m})^{3} \rangle+12\langle \bar{\delta}(r_{m}) \rangle^{2}\langle \bar{\delta}(r_{m})^{2} \rangle \\
&\quad -3\langle \bar{\delta}(r_{m})^{2} \rangle^{2}-6\langle \bar{\delta}(r_{m}) \rangle^{4} \biggr]\numberthis
\end{align*}
The volume-averaged overdensity can be calculated at the linear and quadratic orders using Eq. \eqref{eq:spherical_fourier}, yielding
\begin{equation}\label{eq:del1}
\bar{\delta}^{(1)}(r_{m})=\frac{1}{\sqrt{2\pi^2}}\frac{4}{9}\int_{0}^{\infty} \frac{dk}{k}(kr_{m})^{4}W(kr_{m})\frac{\zeta(k)}{r_{m}^2}
\end{equation}
and
\begin{equation}\label{eq:del2}
\bar{\delta}^{(2)}(r_{m})=-\frac{2}{27}\frac{1}{2\pi^2}\int_{0}^{\infty}\frac{dk_{1}}{k_{1}}\frac{dk_{2}}{k_{2}}(k_{1}r_{m})^{4}(k_{2}r_{m})^{4}W(k_{1}r_{m})W(k_{2}r_{m})\frac{\zeta(k_{1})}{r_{m}^{2}}\frac{\zeta(k_{2})}{r_{m}^{2}}
\end{equation}
The function $W(kr_{m})$ is the top hat window function defined as
\begin{equation}
W(kr_{m})=\frac{3}{(kr_{m})^3}\left( \sin kr_{m}-kr_{m}\cos kr_{m} \right)
\end{equation}
This naturally appears in the calculation by virtue of the spherical decomposition of the curvature perturbations- as opposed to choosing one out of convenience. Another popular choice is the Gaussian window function which has been extensively used in the study of PBH formation, especially because of its nice analytical properties. However, the choice of different cosmological smoothing quantitatively alters the results as pointed out in \cite{Ando:2018qdb}. As it turns out, a Gaussian filter is rather effective at removing small scale (large momentum) fluctuations so that they do not contribute to the final result. On the other hand, a top hat filter is not so efficient and such a smoothing introduces oscillating contributions at small scales. As a result, a subhorizon transfer function is often employed to dampen these small scale oscillatory effects. Nevertheless, non negligible contributions still remain even after the use of an appropriate transfer function. More details on this can also be found in Appendix (\ref{appendix:C}). It will be discussed later in the paper that, due to the behaviour of the top hat filter, numerical results of PBH abundance will be altered in regards to the threshold density $\bar{\delta}_{\text{th}}$ and the peak of the curvature power spectra.\footnote{The altered abundance criteria is different insofar as one compares the results with the Gaussian window function. An important difference arises from relative difference between the smoothed variance produced by these two filters. Readers are referred to \cite{ballesteros,dalianis} where the Gaussian filter have been used.}\\\\
\indent The cumulants are calculated from the expansion of $\bar{\delta}$ upto quadratic order, making use of the definition of $\langle \zeta\zeta \rangle$ from Eq. \eqref{eq:spherical_fourier}. Higher order correlation functions which appear are simplified into products of two point functions using Wick's theorem. Then, the cumulants $\sigma^{2}$, $\tilde{\kappa}_{3}$ and $\tilde{\kappa}_{4}$ are
\begin{align*}
\sigma^{2}(r_{m})&\simeq\langle \bar{\delta}^{(1)}(r_{m})^2 \rangle+\mathcal{O}(\zeta^4) \\ 
&=\frac{16}{81}\int_{0}^{\infty}\frac{dk}{k}(kr_{m})^{4}W(kr_{m})^2T(kr_{m})^2\mathcal{P}_{\zeta}(k) \numberthis
\end{align*}
and
\begin{align*}
\tilde{\kappa}_{3}(r_{m})&\simeq\frac{3}{\sigma(r_{m})^{3}}\left[ \langle \bar{\delta}^{(1)}(r_{m})^{2}\bar{\delta}^{(2)}(r_{m})\rangle-\langle \bar{\delta}^{(1)}(r_{m})^{2} \rangle  \langle \bar{\delta}^{(2)}(r_{m}) \rangle\right]+\mathcal{O}(\zeta^6) \\ 
&=  -\frac{1}{\sigma(r_{m})^{3}}\left(\frac{4}{9}\right)^{3}\int_{0}^{\infty}\frac{dk_{1}}{k_{1}}\frac{dk_{2}}{k_{2}}(k_{1}r_{m})^{4}(k_{2}r_{m})^{4}W(k_{1}r_{m})^{2}W(k_{2}r_{m})^{2} \\
& \qquad \;\;\;\;\;\;\;\;\;\;\;\;\;\;\;\;\;\;\;\;\;\;\;\;\;\;\;\;\;\;\times T(k_{1}r_{m})^{2}T(k_{2}r_{m})^{2}\mathcal{P}_{\zeta}(k_{1})\mathcal{P}_{\zeta}(k_{2}) \\ 
&=-\frac{9}{4}\sigma(r_{m}) \numberthis
\end{align*}
and
\begin{align*}
\tilde{\kappa}_{4}(r_{m})&=\frac{3}{\sigma(r_{m})^{4}}\cdot\left( \frac{4}{9} \right)^{5}\int_{0}^{\infty}\frac{dk_{1}}{k_{1}}\frac{dk_{2}}{k_{2}}\frac{dk_{3}}{k_{3}}(k_{1}r_{m})^{4}(k_{2}r_{m})^{4}(k_{3}r_{m})^{4}W(k_{1}r_{m})^{2}W(k_{2}r_{m})^{2}W(k_{3}r_{m})^{2}\\
&\qquad\;\;\;\;\;\;\;\;\;\;\;\;\;\;\;\;\;\;\;\;\;\;\;\;\;\;\;\;\times T(k_{1}r_{m})^{2}T(k_{2}r_{m})^{2}T(k_{3}r_{m})^{2}\mathcal{P}_{\zeta}(k_{1})\mathcal{P}_{\zeta}(k_{2})\mathcal{P}_{\zeta}(k_{3})\\
&=\frac{27}{4}\sigma(r_{m})^{2}\numberthis
\end{align*}
where $T(kr_{m})$ is the linear transfer function \cite{Mukhanov:2005sc} that has been artificially introduced into the calculation to get a better handle on the oscillatory large momentum modes. Transfer functions describe the evolution of perturbations on subhorizon scales and the one used here is defined for the RD epoch.
\begin{equation}
T(k\eta)=3\frac{\sin c_{s}k\eta - c_{s}k\eta\cos c_{s}k\eta}{(c_{s}k\eta)^3}
\end{equation}
where $c_{s}^{2}=1/3$ is the sound speed of the relativistic fluid and $\eta$ is the conformal time. As a result of the inclusion of the nonlinearities, there is a non negligible contribution coming from the skewness and kurtosis. Qualitatively, a negative skew has the tendency of pushing the PDF along the positive end of the tail of distribution. On the otherhand, a positive kurtosis has the effect of actually increasing the area under the tail of the PDF. The latter has the effect of increasing PBH abundance as will be seen in the next sections.
\section{Non-Gaussian modification of Press-Schechter formalism}
\label{sec:PS_mod}
The fact that there are nontrivial contributions from the third and fourth cumulants present in the statistics of $\bar{\delta}$ tells us that a modification of the PBH abundance is to be expected. The Press-Schechter formalism has been commonly used to assign mass to halos and, as such, has found its way into computing mass fractions of PBHs. The window functions are used to smooth cosmological density fields to some scale which also have characterisic masses associated with them. Then Press and Schechter postulated that the fraction of overdensity fields, averaged over some volume containing mass $M$, larger than some threshold will correspond to the fraction of collapsed objects with mass greater than $M$ \cite{press1974,2010gfe..book.....M}. Then, for Gaussian perturbations, the PBH formation fraction reads 
\begin{equation}
\beta_{\text{G}}(M(k))=\int_{\delta_{\text{th}}}^{\delta_{\text{max}}}\frac{d\delta_{\text{G}}}{\sqrt{2\pi}\sigma(k)}e^{-\delta^{2}_{\text{G}}/2\sigma(k)^2}
\end{equation}
where the variance $\sigma(k)^{2}$ is the same as the one defined in the previous chapter and $\delta_{\text{G}}$ denotes Gaussian overdensity perturbations. The wavenumber $k$ can be related to the horizon mass via \cite{Ozsoy:2018flq}
\begin{equation}\label{eq:PBH_mass}
M(k)\simeq 1.6\times 10^{18}\text{g}\left( \frac{g_{\star}(T_{k})}{106.75} \right)^{-1/6}\times\left( \frac{k}{5.5\times 10^{13}\text{Mpc}^{-1}} \right)^{-2}
\end{equation}
where $g_{\star}(T_{k})$ is the effective number of relativistic degrees of freedom at the time of PBH formation (which is equal to 106.75 during the RD epoch). Implicit in the computations is the assumption that $r_{m}^{-1}\sim k$, which is not entirely accurate but is used regardless due to computational simplicity. The effects of non-Gaussianity can be studied by treating the higher order cumulants as perturbations on the Gaussian distributed overdensity $\delta_{\text{G}}$ \cite{Kawasaki:2019mbl}\footnote{This was similarly considered in \cite{Byrnes:2012yx,Young:2013oia} as a means of including primordial non-Gaussianity by introducing the $f_{\text{NL}}$ and $g_{\text{NL}}$ parameters as an expansion to the Gaussian curvature perturbations such that $$ \zeta =\zeta_{\text{g}}+\frac{3}{5}f_{\text{NL}}(\zeta_{\text{g}}^{2}-\sigma^{2})+\frac{9}{25}g_{\text{NL}}\zeta_{\text{g}}^{3}$$}
\begin{equation}\label{eq:delprime}
\bar{\delta}[\delta_{\text{G}}]=\delta_{\text{G}}+\frac{\tilde{\kappa}_{3}}{6\sigma}\left( \delta_{\text{G}}^{2}-\sigma^2 \right)+\left( \frac{\tilde{\kappa}_{4}}{18\sigma^{2}} \right)^{2}\delta_{\text{G}}^{3}
\end{equation}
such that $\langle \bar{\delta} \rangle=0$. The non-Gaussian modification to the probability density can then be obtained by a formal change of variables
\begin{equation}
P_{\text{NG}}=\sum_{i}\bigg\lvert \frac{d\delta_{\text{G};i}}{d\bar{\delta}} \bigg\lvert P_{\text{G}}[\delta_{\text{G};i}(\bar{\delta})]
\end{equation}
The sum over $i$ corresponds to the roots of the equation $\delta_{\text{G}}(\bar{\delta})=0$. Considering the case where $\tilde{\kappa}_{4}=0$,
\begin{equation}
\delta_{\text{G};i=\pm}=\frac{3\sigma}{\tilde{\kappa}_{3}}\left( -1\pm \sqrt{1+\frac{2\tilde{\kappa}_{3}}{3}\left( \frac{\tilde{\kappa}_{3}}{6}+\frac{\bar{\delta}}{\sigma} \right)} \right)
\end{equation}
Then, the non-Gaussian modification to the PBH formation fraction is given by
\begin{equation}\label{eq:beta_NG}
\beta_{\text{NG}}(M(k))=\int_{\bar{\delta}_{\text{th}}}P_{\text{NG}}(\bar{\delta};k)d\bar{\delta}
\end{equation}
The lower limit of the integral $\bar{\delta}_{\text{th}}$ is the threshold value for the volume-averaged overdensity. As a result, the probability density function is integrated in the range $0.5 \leq \bar{\delta} < 2/3$ according to \cite{Musco:2018rwt}. In effect, the PBHs considered are type I. Similarly, if $\tilde{\kappa}_{3}=0$, the resulting $\delta_{\text{G}}(\bar{\delta})=0$ will possess only one real solution.
\begin{multline*}
\delta_{\text{G}}=6\cdot6^{2/3}\sigma^{4}\left( \bar{\delta}\tilde{\kappa}_{4}^{4}\sigma^{4} \right)^{-1/3}\left( -1+\sqrt{1+\frac{48}{\bar{\delta}^{2}\tilde{\kappa}_{4}^{2}}} \sigma^{4}\right)^{-1/3}\\-\frac{3\cdot 6^{1/3}}{\tilde{\kappa}_{4}^{2}}\left( \bar{\delta}\tilde{\kappa}_{4}^{4}\sigma^{4} \right)^{1/3}\left( -1+\sqrt{1+\frac{48}{\bar{\delta}^{2}\tilde{\kappa}_{4}^{2}}} \sigma^{4}\right)^{1/3}\numberthis
\end{multline*}
The formation fraction, again, can be computed by direct integration as in Eq. \eqref{eq:beta_NG}. Nevertheless, calculations can be simplified if a new variable $y=\frac{\delta_{\text{G}}}{\sigma}$ is introduced. This effectively ensures that the final result can be obtained by integrating Gaussian functions in terms of the new variable and can be expressed using error functions (however, some care must be taken in reestablishing the new expression of $\beta_{\text{NG}}$ in the case of $\tilde{\kappa}_{3}<0$, details of which can be found in \cite{Byrnes:2012yx}). Hence, the final expression for the non-Gaussian formation fraction reads

\[
  \beta_{\text{NG}}(M(k)) =
  \begin{cases}
  \frac{1}{2}\text{erfc}\left( \frac{y_{\text{th}}^{-}}{\sqrt{2}} \right)-\frac{1}{2}\text{erfc}\left( \frac{y_{\text{th}}^{+}}{\sqrt{2}} \right) & \tilde{\kappa}_{3}<0\;\text{and}\;\tilde{\kappa}_{4}=0 \\
  \frac{1}{2}\text{erfc}\left( \frac{y_{\text{th}}}{\sqrt{2}} \right)& \tilde{\kappa}_{3}=0\;\text{and}\;\tilde{\kappa}_{4}>0\numberthis
  \end{cases}
\]
where $y_{\text{th}}^{\pm}$ refer to the threshold obtained for the two roots of the nonvanishing skewness case. Since $\beta\equiv\rho_{\text{PBH}}/\rho_{\text{rad}}$, the resulting abundance grows proportional to the scale factor $a$ during RD, only reaching a constant value at the onset of matter domination, at which point the fraction of PBH over CDM is determined
\begin{equation}
f_{\text{PBH}}(M)=\frac{1}{\Omega_{\text{CDM}}}\frac{d\Omega_{\text{PBH}}}{d\ln M}=\left(\frac{\beta(M)}{8\times10^{-15}}\right)\left( \frac{\Omega_{\text{CDM}}h^2}{0.12} \right)^{-1}\left( \frac{g_{\star}}{106.75} \right)^{-1/4}\left( \frac{M}{M_{\odot}} \right)^{-1/2}
\end{equation}
Here $\Omega_{\text{CDM}}h^2 =0.12$ is the current CDM energy density \cite{Aghanim:2018eyx}.

\section{PBH abundance calculation}
\label{sec:PBH_abundance}
In this section, the PBH abundance will be calculated using both the Gaussian and non-Gaussian expressions for PBHs in the mass range $10^{17}-10^{18}\text{g}$. Such a mass range is of interest in cosmology since observational constraints for PBH dark matter is potentially unconstrained (readers are referred to \cite{Carr:2020gox} for updated constraints on PBH dark matter). The types of curvature power spectrum which can be considered are (i) one with a central spike (Dirac delta function) and (ii) with finite width (lognormal). The lognormal function serves as a more accurate functional representation of $\mathcal{P}_{\zeta}$ around the peak where a final calculation can be carried out using results from a concrete inflation model which predicts a peaked curvature power spectrum at a certain scale. Although femtolensing constraints on PBH CDM have been relaxed recently in this mass range \cite{Katz:2018zrn}, the calculations will be restricted to $f_{\text{PBH}}\sim 0.1$ which can be obtained when $\beta\sim 10^{-17}.$ Also, using the correspondance between the compaction function and the volume-averaged overdensities, the collapse threshold is taken to be $\bar{\delta}_{\text{th}}\sim 0.55$.
\subsection{Dirac delta function $\mathcal{P}_{\zeta}$} 
\begin{figure}
\includegraphics[scale=0.55]{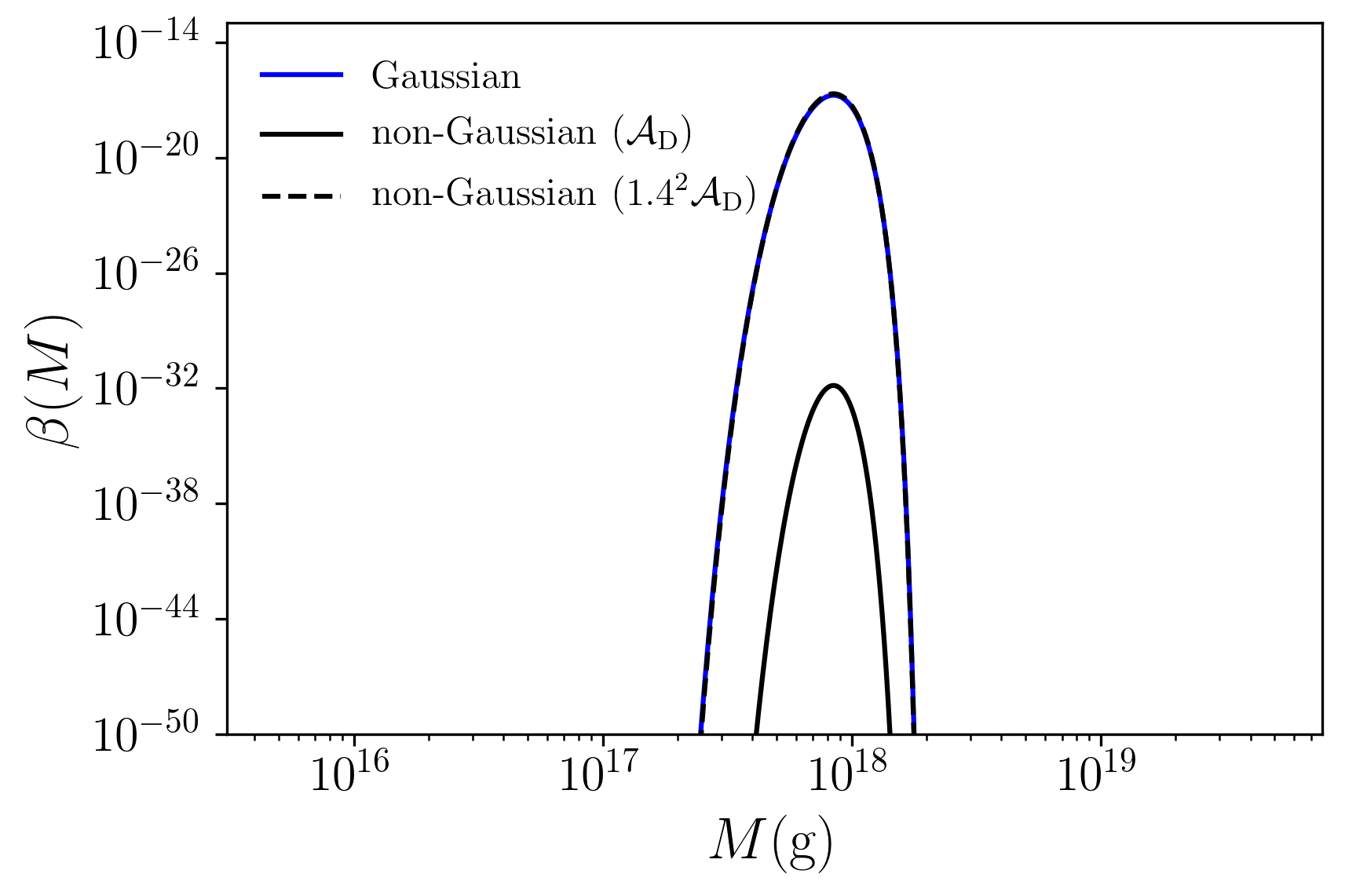}
\includegraphics[scale=0.55]{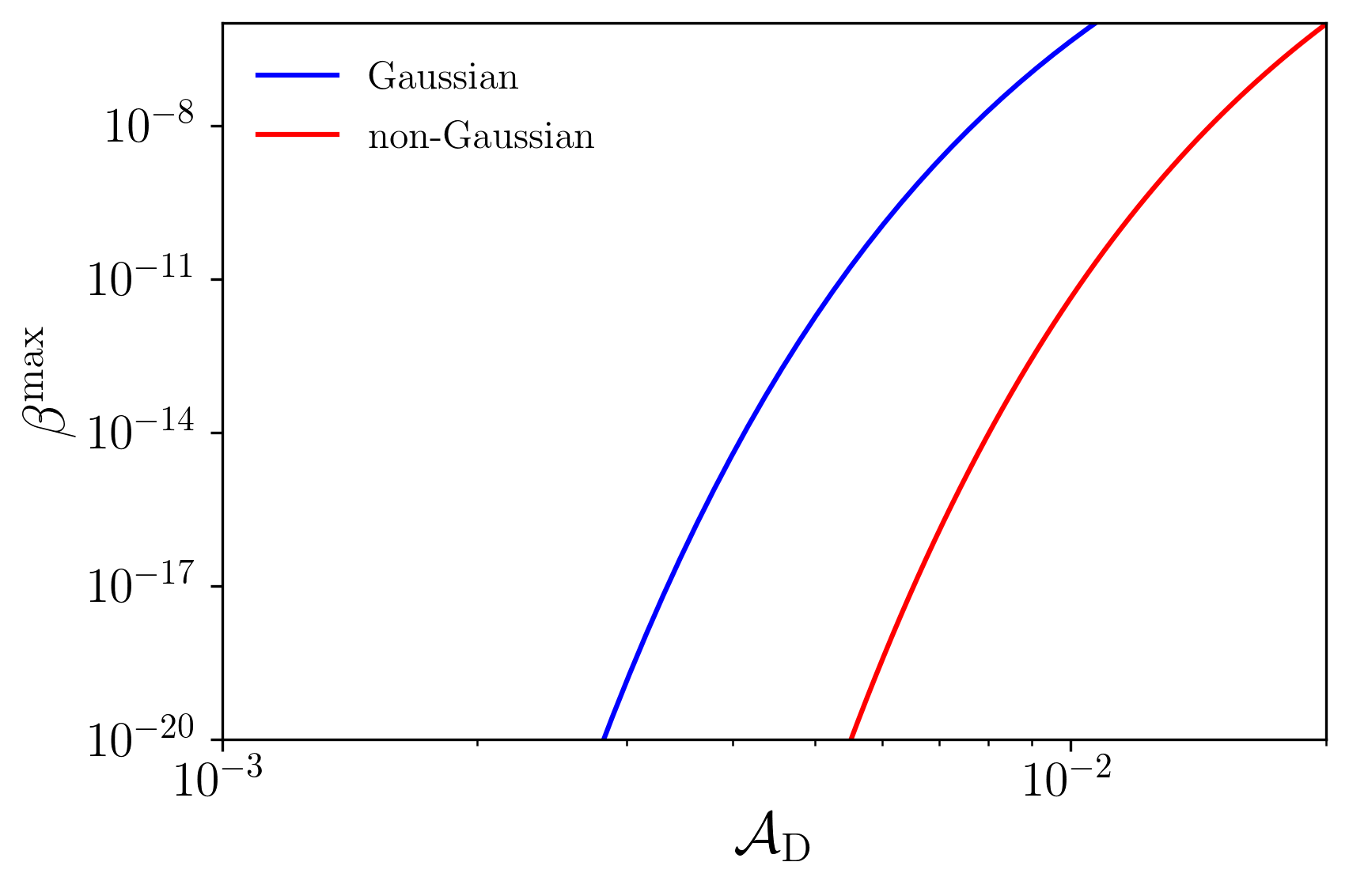}
\caption{\textit{Left panel}: PBH abundance $\beta(M)$ at formation for Gaussian overdensities and non-Gaussian overdensities with two different amplitudes for Dirac delta $\mathcal{P}_{\zeta}$ with $\tilde{\kappa}_{3}<0$ and $\tilde{\kappa}_{4}=0$; \textit{Right panel}: Variation of the peak value of abundance $\beta^{\text{max}}$ with amplitude}
\label{fig:dirac_delta_stuff}
\end{figure}

\begin{figure}
\centering
\includegraphics[scale=0.55]{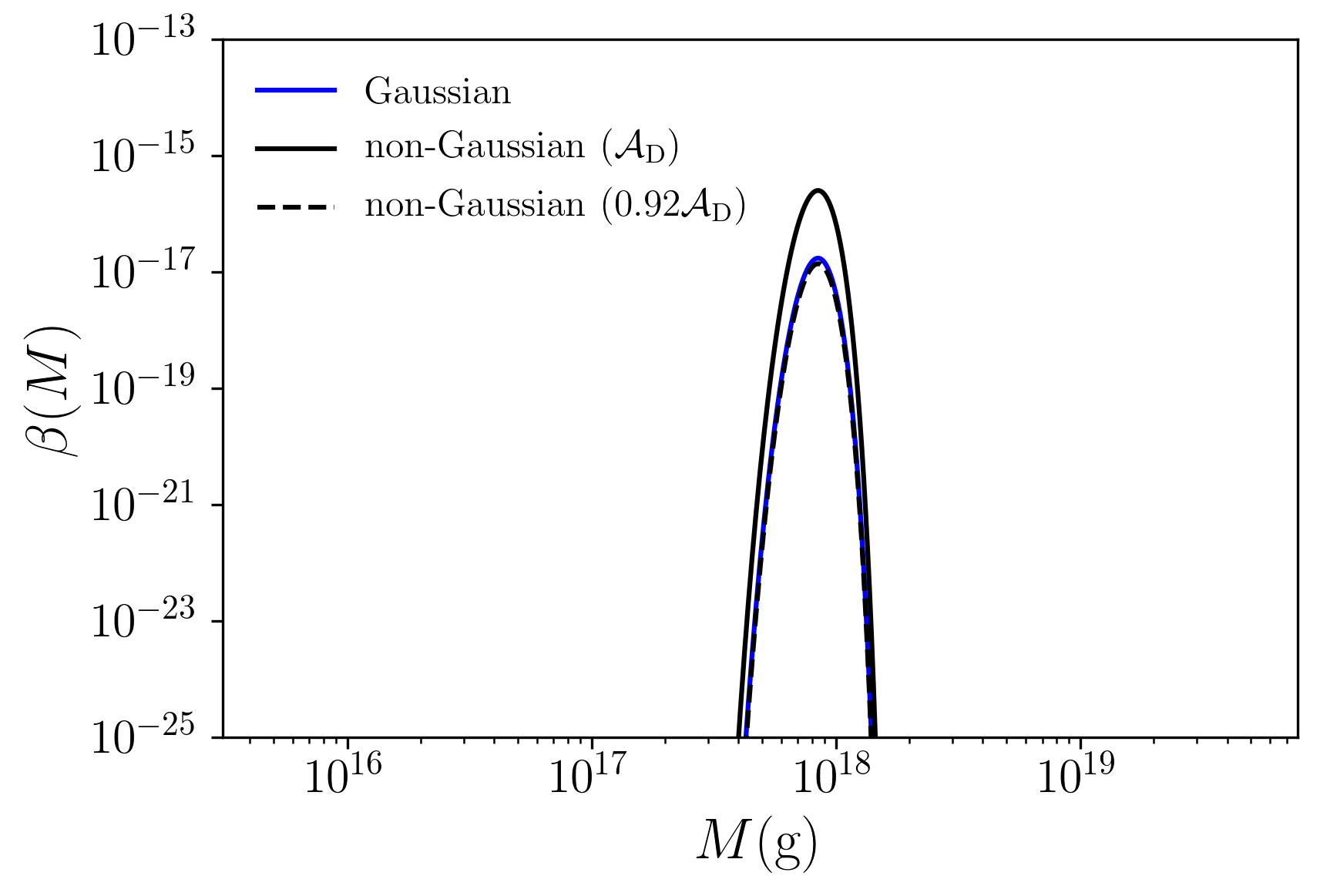}
\caption{PBH abundance $\beta(M)$ at formation for Gaussian overdensities and non-Gaussian overdensities with two different amplitudes for Dirac delta $\mathcal{P}_{\zeta}$ with $\tilde{\kappa}_{3}=0$ and $\tilde{\kappa}_{4}>0$}
\label{fig:dirac_delta_stuff2}
\end{figure}

The Dirac delta function would correspond to a sharp peak in the inflationary power spectrum. Although it is rather unphysical since USR infation models produce peaked curvature power spectra of finite width. Then, the power spectrum could potentially take the following form
\begin{equation}\label{eq:delta_power}
\mathcal{P}_{\zeta}(k)=\mathcal{A}_{\text{D}}k_{\star}\delta(k-k_{\star})
\end{equation}
where $k_{\star}$ is the characteristic scale at which the power spectrum peaks and $\mathcal{A}_{\text{D}}$ is a parameter that controls the amplitude of $\mathcal{P}_{\zeta}$. To produce PBHs in the mass range $10^{17}-10^{18}\text{g}$, the characteristic scale is chosen to be $k_{\star}\sim 1.9\times 10^{14}\text{Mpc}^{-1}$. The results for the Dirac delta function power spectrum have been summarized in Fig. (\ref{fig:dirac_delta_stuff}) for the case of $\tilde{\kappa}_{3}<0$ and $\tilde{\kappa}_{4}=0$. It can be seen that the negative skewness suppresses the abundance by many orders of magnitude. The non-Gaussian abundance becomes comparable to that of the Gaussian one when the amplitude is modified by $1.4^{2}\mathcal{A}_{\text{D}}$.\\\\
\indent On the other hand, if only the kurtosis is considered ($\tilde{\kappa}_{3}=0$ and $\tilde{\kappa}_{4}>0$), shown in Fig. (\ref{fig:dirac_delta_stuff2}), the effect of this mode of non-Gaussianity is not as severe as the negative skewness case and the non-Gaussian abundance becomes comparable to the Gaussian one when the amplitude is modified to $0.92\mathcal{A}_{\text{D}}$, which is a relatively minor adjustment. The $\beta(M)$ plots for both skewness and kurtosis cases in Fig. (\ref{fig:dirac_delta_stuff}) and (\ref{fig:dirac_delta_stuff2}) have been created for $\mathcal{A}_{\text{D}}=0.0034$.

\subsection{Lognormal $\mathcal{P}_{\zeta}$}\label{sec:lognormal}
\begin{figure}
\includegraphics[scale=0.55]{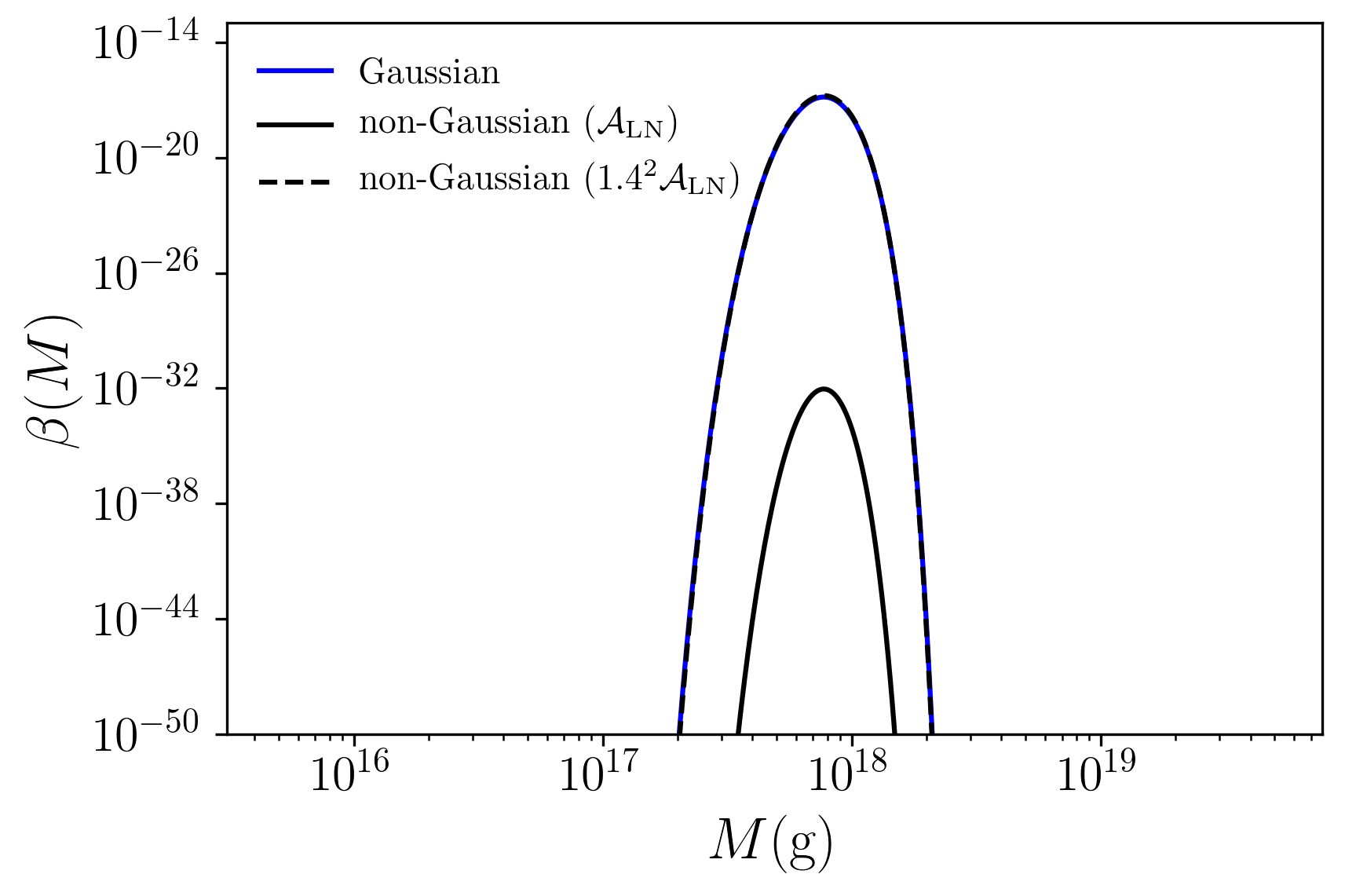}
\includegraphics[scale=0.55]{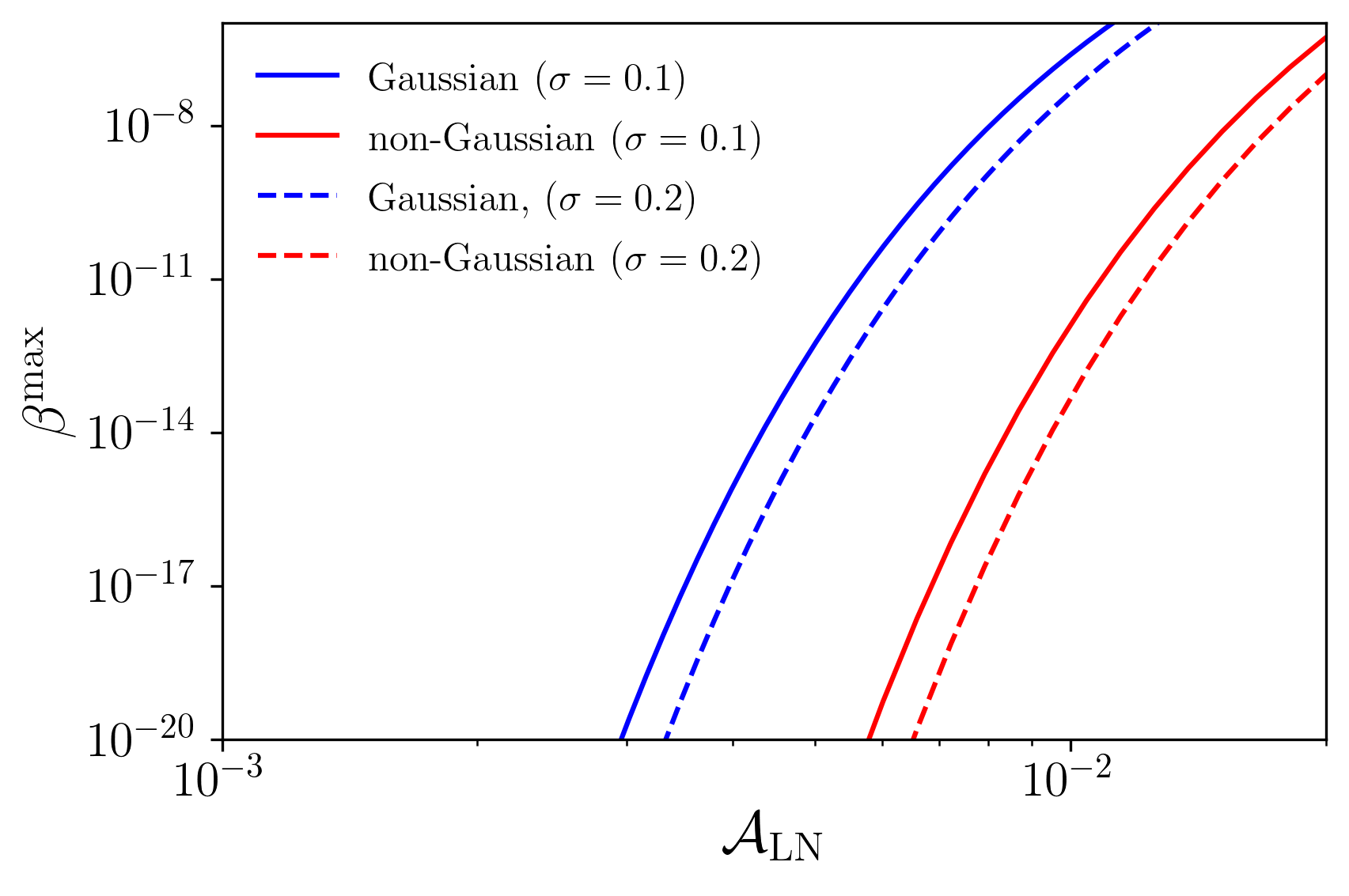}
\caption{\textit{Left panel}: PBH abundance $\beta(M)$ at formation for Gaussian overdensities and non-Gaussian overdensities with two different amplitudes for lognormal $\mathcal{P}_{\zeta}$ with $\tilde{\kappa}_{3}<0$ and $\tilde{\kappa}_{4}=0$; \textit{Right panel}: Variation of the peak value of abundance $\beta^{\text{max}}$ with amplitude}
\label{fig:lognormal_stuff}
\end{figure}

\begin{figure}
\centering
\includegraphics[scale=0.55]{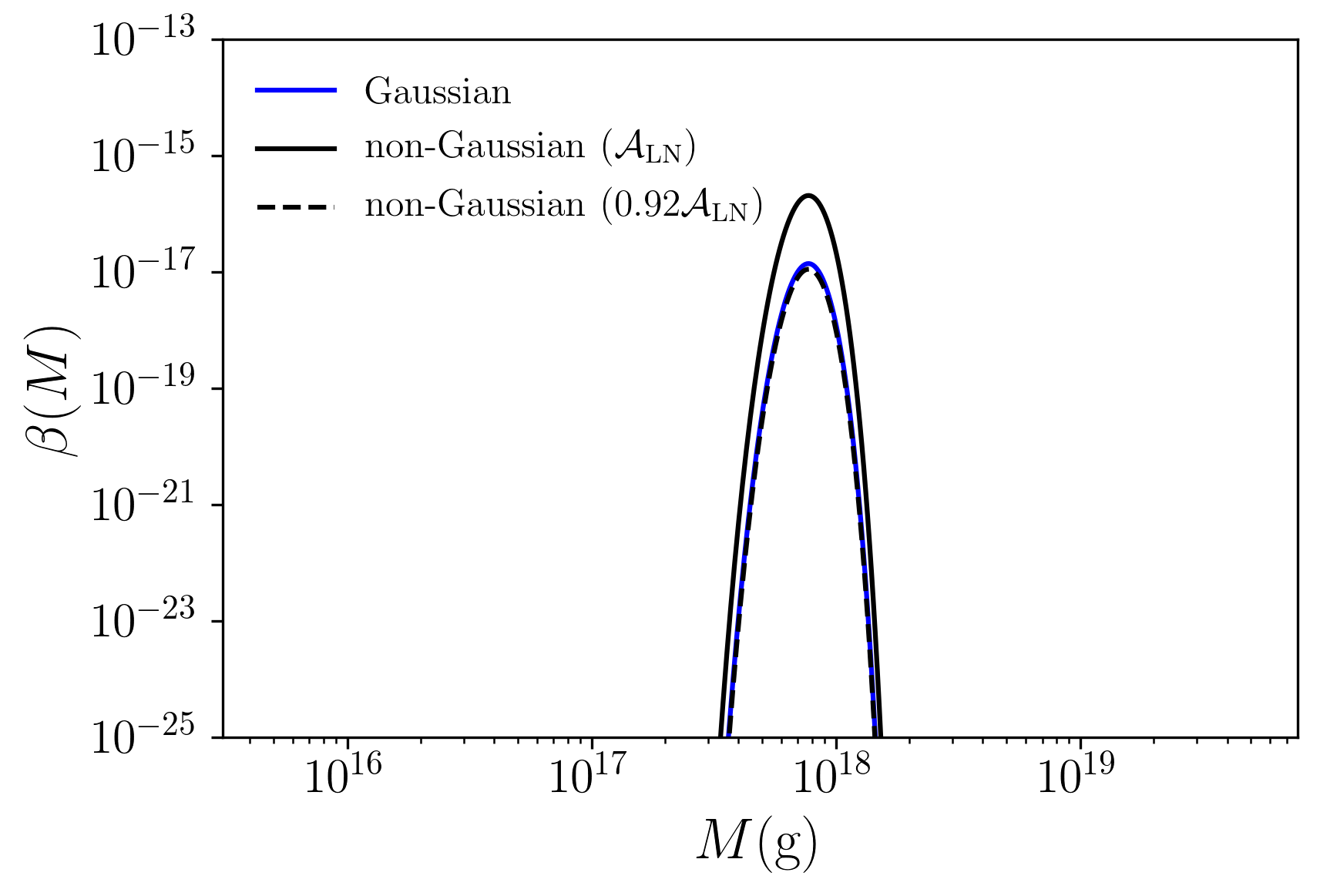}
\caption{PBH abundance $\beta(M)$ at formation for Gaussian overdensities and non-Gaussian overdensities with two different amplitudes for lognormal $\mathcal{P}_{\zeta}$ with $\tilde{\kappa}_{3}=0$ and $\tilde{\kappa}_{4}>0$}
\label{fig:lognormal_stuff2}
\end{figure}

The USR inflation models that have recently gained popularity in the study of PBH formation usually predict peaked curvature power spectra where the peak can be modelled using a lognormal distribution of the form
\begin{equation}\label{eq:lognormal}
\mathcal{P}_{\zeta}(k)=\frac{\mathcal{A}_{\text{LN}}}{\sqrt{2\pi\sigma^{2}}}\exp\left( -\frac{\ln^{2}(k/k_{\star})}{2\sigma^{2}} \right)
\end{equation}
Here the distribution is defined by an additional parameter $\sigma$ which helps control the width of the power spectrum in question, while $\mathcal{A}_{\text{LN}}$ controls the amplitude. Not unlike the Dirac delta function $\mathcal{P}_{\zeta}$, the characteristic scale here is also $k_{\star}\sim 1.9\times 10^{14}\text{Mpc}^{-1}$ with the results summarized in Fig. (\ref{fig:lognormal_stuff}) and (\ref{fig:lognormal_stuff2}). The results are very similar to those of the Dirac delta function power spectrum where, in the $\tilde{\kappa}_{3}<0$ case, the non-Gaussian abundance becomes comparable to the Gaussian one when the amplitude is modified by $1.4^{2}\mathcal{A}_{\text{LN}}$. We see, from the plot on the right in Fig. (\ref{fig:lognormal_stuff}), that the general tendency of the peak value of the formation fraction is to shift towards higher values of $\mathcal{A}_{\text{LN}}$. Hence a more realistic $\mathcal{P}_{\zeta}$ with a larger width would require a higher amplitude in order to reach $\beta\sim 10^{-17}$. It is also worthwhile to note that the shape of the $\beta^{\text{max}}$ curve for $\sigma=0.1$ is most similar to the Dirac delta function case and it would be interesting to observe whether these two approach each other in the very small $\sigma$ limit.\\\\
\indent The case of the nonvanishing kurtosis is also similar to that of the Dirac delta function $\mathcal{P}_{\zeta}$ as seen in Fig. (\ref{fig:lognormal_stuff2}) where the non-Gaussian abundance becomes comparable to the Gaussian one when the amplitude is modified to $0.92\mathcal{A}_{\text{LN}}$. The $\beta(M)$ plots for both skewness and kurtosis cases in Fig. (\ref{fig:lognormal_stuff}) and (\ref{fig:lognormal_stuff2}) have been created for $\mathcal{A}_{\text{LN}}=0.004$ and $\sigma=0.2$.

\subsection{Peaked $\mathcal{P}_{\zeta}$ from $\alpha-$attractors}
\begin{figure}
\centering
\includegraphics[scale=0.55]{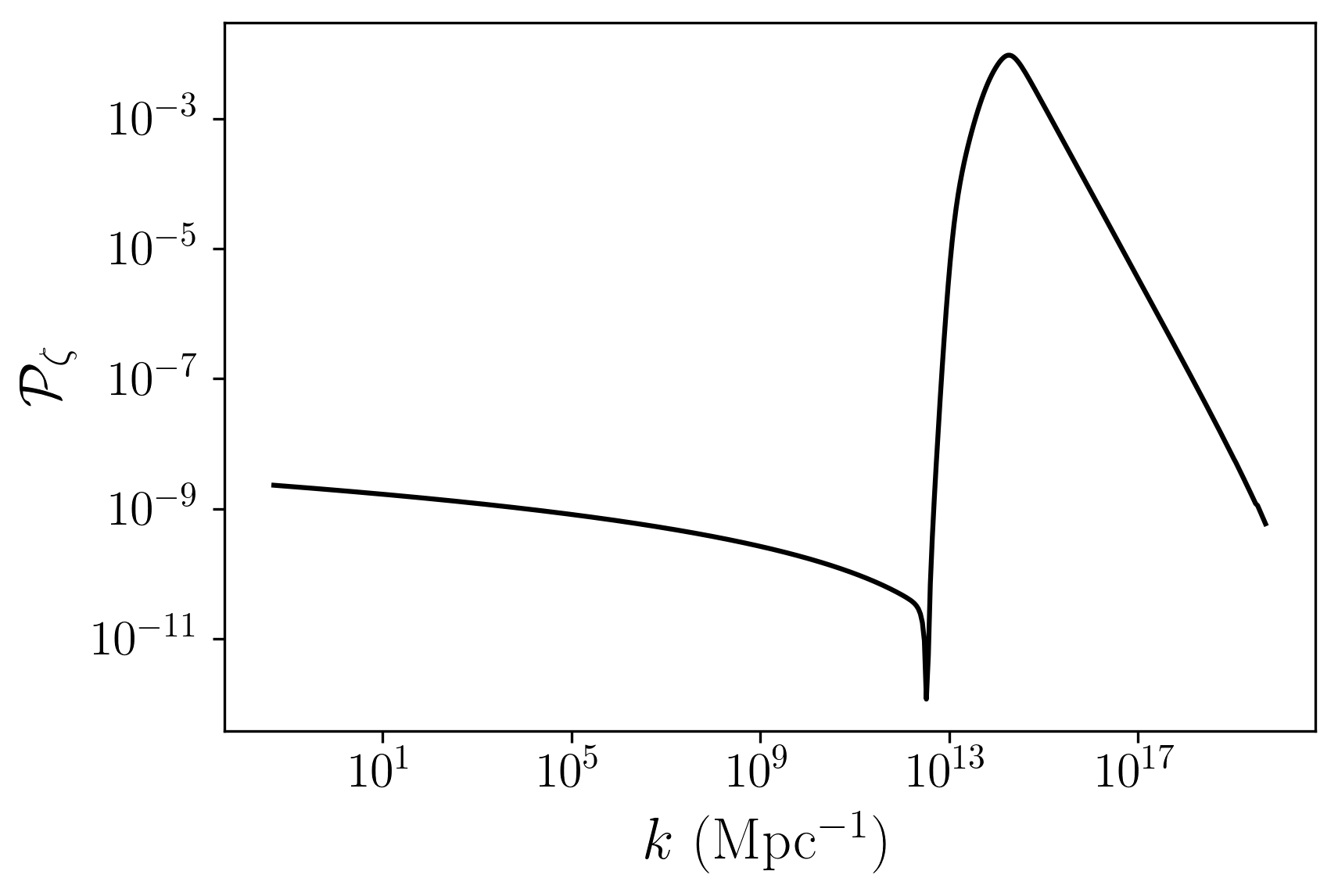}
\caption{Curvature power spectrum obtain from inflationary $\alpha-$attractors}
\label{fig:PS_alpha}
\end{figure}

\begin{figure}
\centering
\includegraphics[scale=0.55]{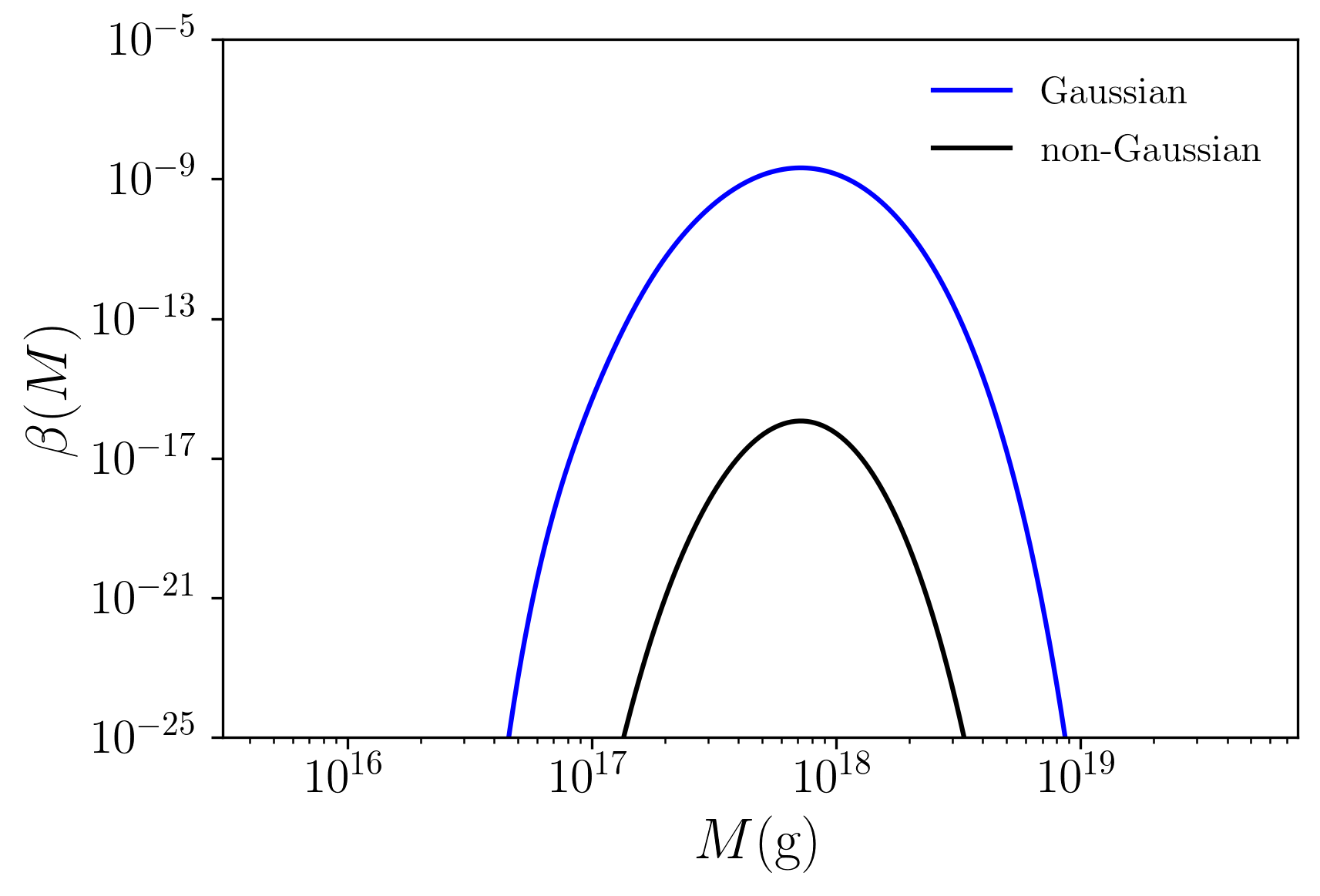}
\includegraphics[scale=0.55]{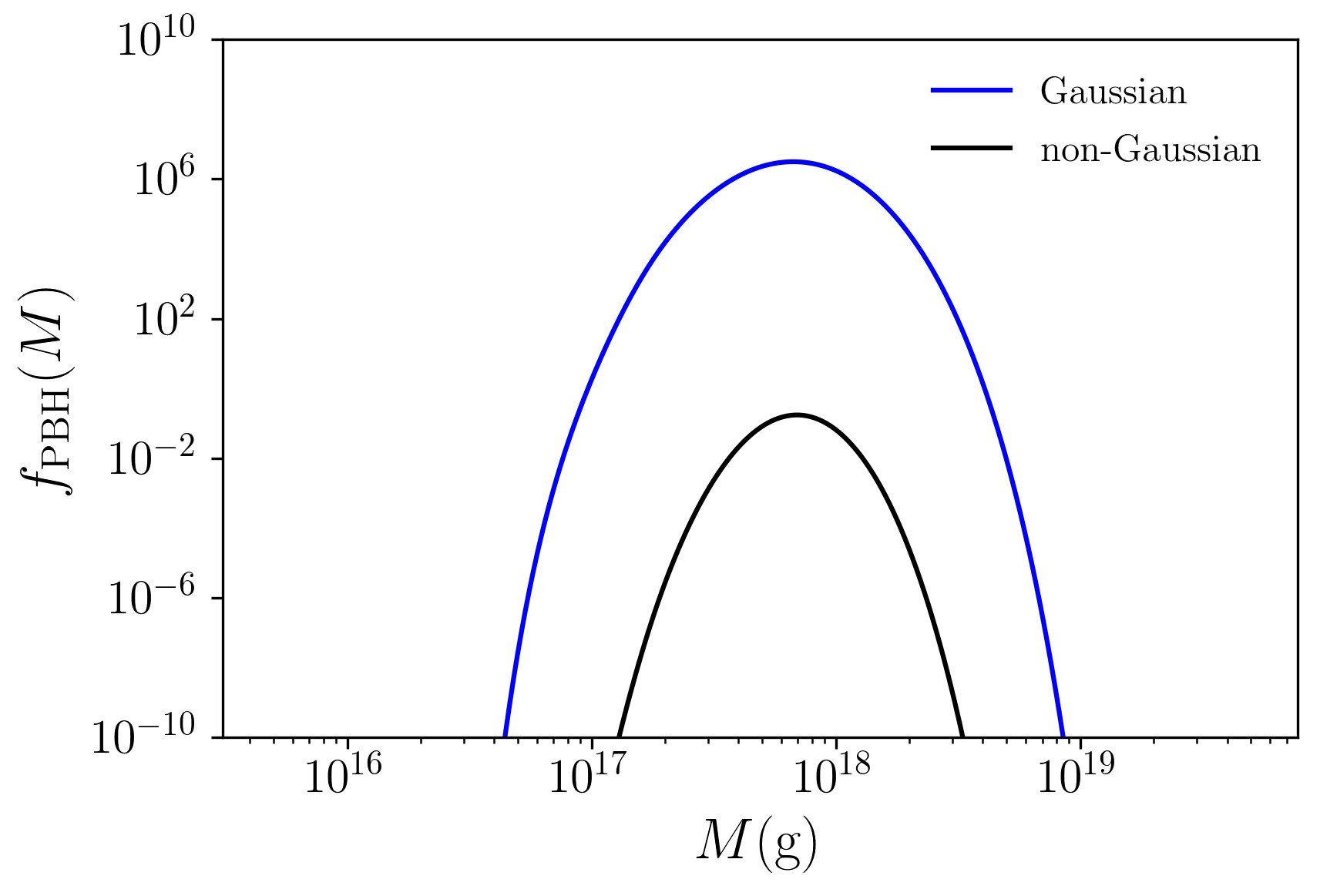}
\caption{\textit{Left panel}: PBH abundance $\beta(M)$ at formation for Gaussian and non-Gaussian overdensities with $\tilde{\kappa}_{3}<0$ and $\tilde{\kappa}_{4}=0$; \textit{Right panel}: The corresponding PBH fraction over CDM}
\label{fig:alpha_attractor_stuff}
\end{figure}

\begin{figure}
\centering
\includegraphics[scale=0.55]{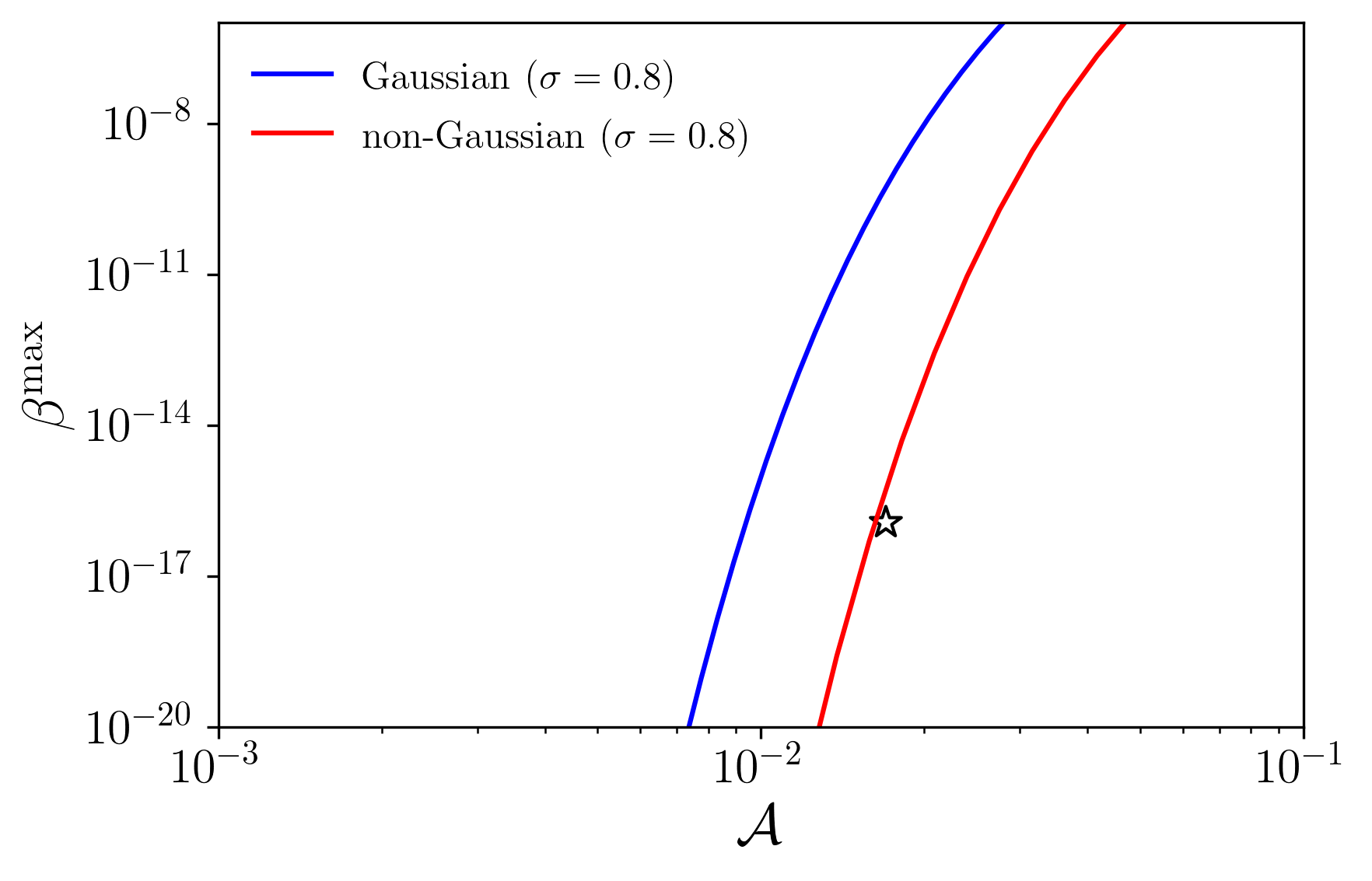}
\caption{Variation of the peak value of abundance $\beta^{\text{max}}$ with amplitude for the $\alpha-$attractor power spectrum fitted with $(\mathcal{A}_{\text{LN}},\sigma)=(0.018,0.8)$}
\label{fig:alpha_attractor_stuff2}
\end{figure}

The inflarionary $\alpha-$attractors have been previously used in the study of PBH formation since it predicts a potential of the form $V(\varphi)=f(\tanh\frac{\varphi}{\sqrt{6\alpha}})^{2}$, where $f$ is an arbitrary holonomic function \cite{Kallosh:2013yoa,Kallosh:2014ona}. This property provides a great deal of freedom in the construction of an inflaton potential, which can be customized into one with an inflection point and hence a USR region. To make a connection with a concrete inflation model, the potential in \cite{Mahbub:2019uhl} will be used, which reads
\begin{equation}
V(\varphi)=V_{0}\left[ 1+a_{1}-\exp\left( -a_{2}\tanh\frac{\varphi}{\sqrt{6\alpha}} \right)-a_{1}\exp\left( -a_{3}\tanh^{2}\frac{\varphi}{\sqrt{6\alpha}} \right) \right]^{2}
\end{equation}
The curvature power spectrum that is produced from such a model is shown in Fig. (\ref{fig:PS_alpha}) from parameter set 2 in \cite{Mahbub:2019uhl}, which peaks at $\mathcal{P}_{\zeta}\sim 9\times 10^{-3}$. The Fourier mode which results in the peak power spectrum is $k=1.9\times 10^{14}\text{Mpc}^{-1}$ the reason for which it serves as the characteristic scales in the toy model power spectra. Although this $\mathcal{P}_{\zeta}$ looks rather complicated, the relevant range of $k-$modes around the peak can be reliably modelled using a lognormal function much like Eq. \eqref{eq:lognormal} with $\mathcal{A}_{\text{LN}}\simeq 0.018$ and $\sigma\simeq 0.8$. Using these, the abundance $\beta(M)$ and the fraction of PBH over CDM $f_{\text{PBH}}(M)$ have been plotted in Fig. (\ref{fig:alpha_attractor_stuff}). The variation of $\beta^{\text{max}}$ as a function of the amplitude is shown in Fig. (\ref{fig:alpha_attractor_stuff2}) where the corresponging non-Gaussian result is shown as the star marker. It can be seen that Gaussian overdensities oversaturate the $f_{\text{PBH}}$ parameter, leading to an overproduction of PBHs. In fact, the energy density of PBHs would be orders of magnitude greater than the currently observed CDM energy density $\Omega_{\text{PBH}}\gg\Omega_{\text{CDM}}$, meaning different parameter sets need to be explored if perturbations are Gaussian. On the other hand, non-Gaussian perturbations fare better and $f_{\text{PBH}}^{\text{max}}\sim 0.17$. This is a reasonable number, even if the femtolensing constraint coming from gamma-ray bursts in this mass range is not relaxed. The positive kurtosis case is not explored here since the abundance remains more or less close to the Gaussian abundance.

\section{Comparison with peaks theory}\label{sec:peaks_compare}

Recently PBH abundance calculation using the theory of peaks has gained a lot of attention \cite{Bardeen:1985tr}. Peaks theory was developed under a more rigorous mathematical setting and has the added advantage that the characteristic scale of the peaks is build into the theory and there is no need for the introduction of a window function, unlike Press-Schechter. Peaks theory has previously been used in \cite{DeLuca:2019qsy,Germani:2018jgr} in the context of PBH formation, however, an optimized criterion for computing PBH formation fraction based on peaks was developed in \cite{Yoo:2018kvb}. Their techniques will be used to compute $\beta(M)$ for Dirac delta and extended power spectra, which will then be compared with the Press-Schechter results from the previous section (for detailed derivations, refer to \cite{Yoo:2018kvb} and references therein).\\\\
It is conventional in peaks theory to define the moments of the curvature power spectrum as follows
\begin{equation}\label{eq:sigma_moments}
\sigma_{n}^{2}=\int\frac{dk}{k}k^{2n}\mathcal{P}_{\zeta}(k)
\end{equation}
Considering high peaks centered at the origin, there are two other quantities that characterize the shape of the peaks \cite{Bardeen:1985tr}
\begin{align*}
\mu &= -\zeta(r=0)\\
k_{*}^{2}&=\frac{1}{\mu}\nabla^{2}\zeta(r)|_{r=0}\numberthis
\end{align*}
where $\mu$ and $1/k_{*}$ define the amplitude and curvature scale of perturbations. For a given form of the curvature perturbation $\zeta(r)$, one may use $\mu$ and $k_{*}$ to define a typical curvature profile define by the following
\begin{equation}
\frac{\bar{\zeta}(r)}{\mu}=g(r;k_{*})=g_{0}(r)+k_{*}^{2}g_{1}(r)
\end{equation}
where 
\begin{align*}
g_{0}(r)&=-\frac{1}{1-\gamma^{2}}\left( \psi+\frac{1}{3}R_{*}^{2}\nabla^{2}\psi \right)\\
g_{1}(r)&=\frac{1}{\gamma(1-\gamma^{2})}\frac{\sigma_{0}}{\sigma_{2}}\left( \gamma^{2}\psi+\frac{1}{3}R_{*}^{2}\nabla^{2}\psi \right)\numberthis
\end{align*}
with $\gamma=\sigma_{1}^{2}/(\sigma_{0}\sigma_{2})$, $R_{*}=\sqrt{3}\sigma_{1}/\sigma_{2}$. The function $\psi(r)$ is the normalized two-point function of curvature perturbations, defined as 
\begin{equation}
\psi(r)=\frac{1}{\sigma_{0}^{2}}\int\frac{dk}{k}\frac{\sin kr}{kr}\mathcal{P}_{\zeta}(k)
\end{equation}
Once a model for the power spectrum has been specified, the function $g(r;k_{*})$ can be used to compute the form of the compaction function by substituting $\bar{\zeta}$ into Eq. \eqref{eq:compaction}, from which the scale of the perturbation $r_{m}$, at which the compaction function is maximized, can be determined.\\\\
The abundance of primordial black holes from peaks theory can be calculated, although not in as straightforward a manner as Press-Schechter. The task lies in computing the number of extrema of $\zeta$ in a comoving volume, applying constraints and then extracting the expected number of peaks $n_{\text{pk}}$. As a function $\mu$ and $M$,
\begin{equation}\label{eq:n_peak}
n_{\text{pk}}(\mu,M)d\mu dM=\left( \frac{3}{2\pi} \right)^{3/2}\frac{\sigma_{2}^{2}}{\sigma_{0}\sigma_{1}^{3}}\mu k_{*}f\left( \frac{\mu k_{*}^{2}}{2} \right)P_{1}\left( \frac{\mu}{\sigma_{0}},\frac{\mu k_{*}^{2}}{\sigma_{2}} \right)\bigg\lvert \frac{d\ln \bar{r}_{m}}{dk_{*}}-\mu\frac{dg_{m}}{dk_{*}} \bigg\lvert^{-1}d\mu d\ln M
\end{equation}
where $\bar{r}_{m}$ is the value of $r_{m}$ when $\zeta=\bar{\zeta}$ and $g_{m}=g(\bar{r}_{m};k_{*})$. The number density of PBHs $n_{\text{PBH}}$ can be obtained by integrating Eq. \eqref{eq:n_peak} over $\mu$. Then, the PBH abundance can be expressed as
\begin{align*}\label{eq:beta_peaks}
\beta(M)d\ln M&=\frac{Mn_{\text{PBH}}}{\rho_{\text{rad}}a^{3}}d\ln M \\
&=2\alpha\left( \frac{3}{2\pi} \right)^{1/2}k_{\text{eq}}^{-3}\frac{\sigma_{2}^{2}}{\sigma_{0}\sigma_{1}^{3}}\left( \frac{M}{M_{\text{eq}}} \right)^{3/2}\int_{\mu_{b}}^{\infty}d\mu\;\mu k_{*}f\left( \frac{\mu k_{*}^{2}}{\sigma_{2}} \right)P_{1}\left( \frac{\mu}{\sigma_{0}}, \frac{\mu k_{*}^{2}}{\sigma_{2}}\right)\\
&\qquad \;\;\;\;\;\;\;\;\;\;\;\;\;\;\;\;\;\;\;\;\;\;\;\;\;\;\;\;\;\;\;\;\;\;\;\;\times \bigg\lvert \frac{d\ln \bar{r}_{m}}{dk_{*}}-\mu\frac{dg_{m}}{dk_{*}} \bigg\lvert^{-1}d\ln M \numberthis
\end{align*}
where $\mu_{b}$ is the minimum value that can be taken by $\mu$, which depends on the functional form of $\mu$ and $k_{*}$. At this point, some comments are needed to made regarding the computation of the PBH mass. Usually, the mass of PBHs are assigned as the mass related to the horizon when a particular mode $k$ becomes of the order of the horizon size, given as $k=aH$. The horizon crossing criterion is a bit more subtle than that and, as discussed in \cite{Yoo:2018kvb}, can be expressed as 
\begin{equation}
aH=\frac{a}{R(\bar{r}_{m})}=\frac{1}{\bar{r}_{m}}e^{\mu g_{m}}
\end{equation}
The PBH mass corresponding to such a horizon crossing condition can be expressed as 
\begin{equation}\label{eq:new_PBH_mass}
M=M_{\text{eq}}k_{\text{eq}}^{2}\bar{r}_{m}^{2}e^{-2\mu g_{m}}
\end{equation}
where $M_{\text{eq}}$ and $k_{\text{eq}}$ are the horizon mass and comoving wavenumber at matter-radiation equality. Eq. \eqref{eq:new_PBH_mass} can also be inverted to obtain $\mu=\mu(M,k_{*})$. The consequence of this modified horizon crossing condition will be shown as a shift in the mass where the peak in $\beta$ occurs.\footnote{The shift will be towards higher mass compared to Eq. \eqref{eq:new_PBH_mass}. This will have implications for inflation model building which can produce PBHs.}
\subsection{Dirac delta function $\mathcal{P}_{\zeta}$}
\begin{figure}
\centering
\includegraphics[scale=0.55]{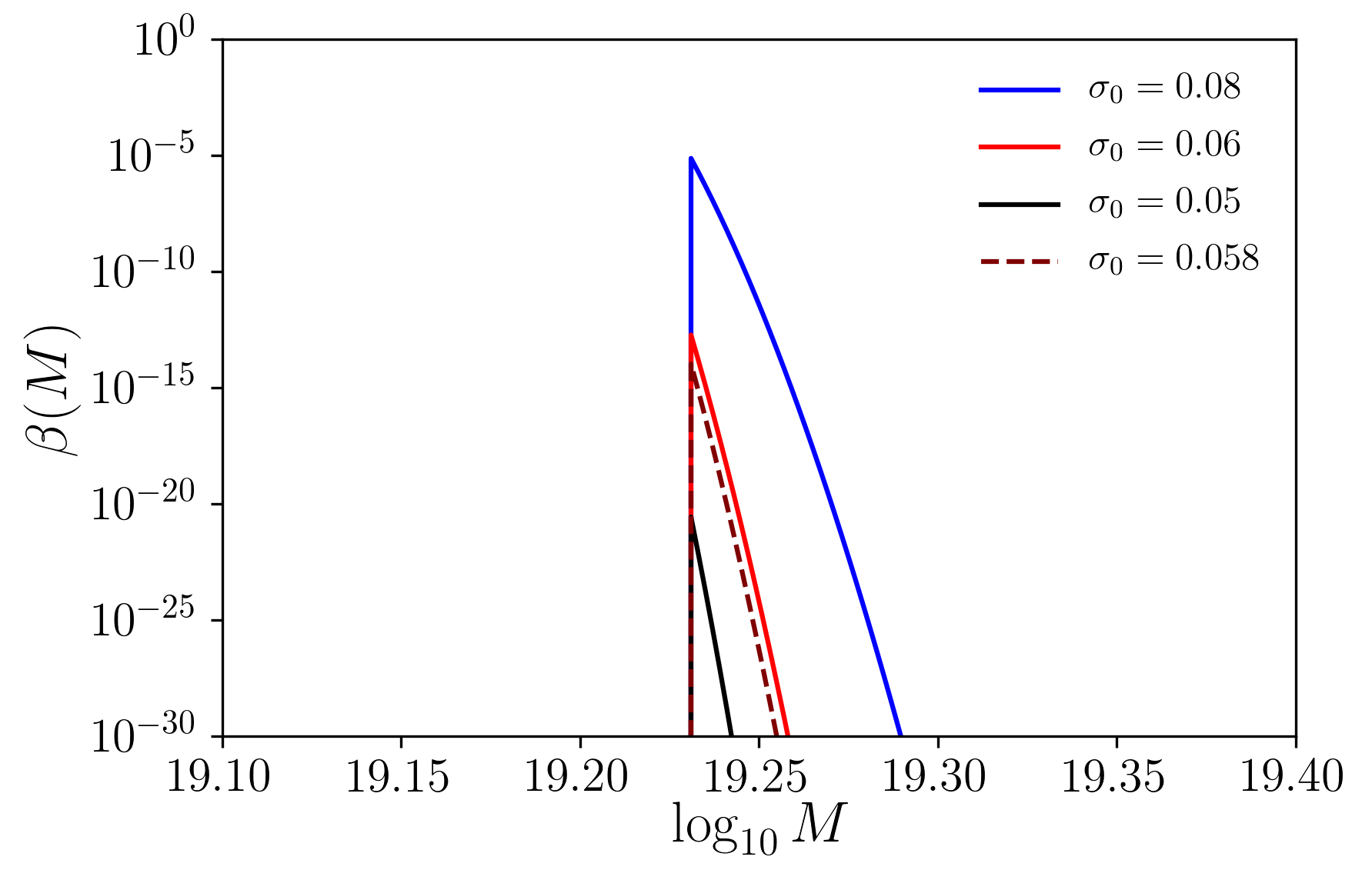}
\includegraphics[scale=0.55]{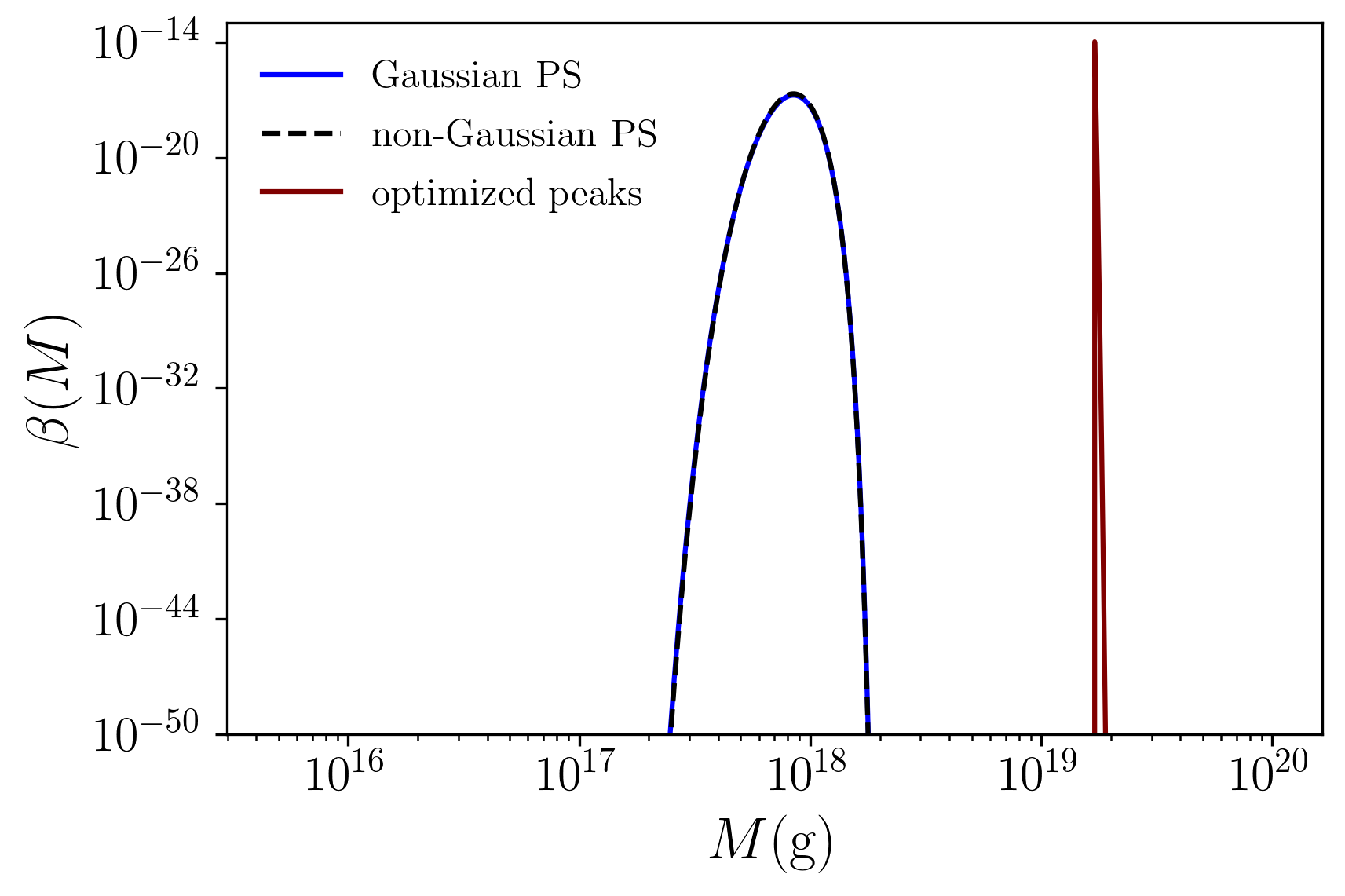}
\caption{\textit{Left panel}: PBH abundance using peaks theory for delta function $\mathcal{P}_{\zeta}$, given by Eq. {\eqref{eq:peaks_beta_dd}} for different amplitudes $\mathcal{A}_{\text{D}}=\sigma_{0}^{2}$; \textit{Right panel}: Comparison with Press-Schechter for $\mathcal{P}_{\zeta}$ with the same amplitude}
\label{fig:peaks_dd_stuff}
\end{figure}

Let us consider the Dirac delta power spectrum given by Eq. \eqref{eq:delta_power}. Using Eq. \eqref{eq:sigma_moments}, it can be shown that the amplitude of the power spectrum is related to the 0-th moment via $\mathcal{A}_{\text{D}}=\sigma_{0}^{2}$, while the $n$-th moments are given as $\sigma_{n}^{2}=\sigma_{0}^{2}k_{\star}^{2n}$. For this power spectrum, it can be easily shown that $k_{c}=\sigma_{1}/\sigma_{0}=k_{\star}$ and $\gamma=1$. The curvature perturbation profile takes the form
\begin{equation}
g(r;k_{\star})=-\psi(r)=-\frac{\sin k_{\star}r}{k_{\star}r}
\end{equation}
Using Eq. \eqref{eq:compaction}, $\mathcal{C}_{\text{max}}$ occurs at $k_{\star}\bar{r}_{m}=l^{2}\simeq 2.74$. Given that the threshold overdensity is $\bar{\delta}_{\text{th}}\sim 0.55$, the threshold value for $\mu$ turns out to be $\mu_{c}\simeq 0.547$, while $g_{c}=-0.141$. The expression for $\beta$ for the case of a spiky power spectrum can be derived following the procedure discussed in \cite{Yoo:2018kvb}, which is simply quoted here
\begin{equation}\label{eq:peaks_beta_dd}
\beta(M)=\frac{3^{1/2}}{2\pi}\frac{1}{\sigma_{0}|g_{c}|}\left( \frac{M}{M_{\text{eq}}} \right)^{3/2}\left( \frac{k_{\star}}{k_{\text{eq}}} \right)^{3}f\left( \frac{\mu_{\star}}{\sigma_{0}} \right)e^{-\mu_{\star}^{2}/2\sigma_{0}^{2}}\Theta(M-M_{c})
\end{equation}
It is worthwhile to take note of the fact that a Heaviside step function appears in the expression for $\beta$, which sets a threshold mass cut-off at $M=M_{c}$. The formation fraction is illustrated in Fig. (\ref{fig:peaks_dd_stuff}). The left panel shows $\beta$ for different amplitudes ($\mathcal{A}=\sigma_{0}^{2}$) where $\sigma_{0}=0.058$ corresponds to the amplitude set in the previous section. From the comparsion with Press-Schechter (as shown in the right panel), we see that the peaks calculation reveals an order of magnitude difference while, at the same time, the PBHs are clustered around the threshold mass $M_{c}\simeq \unit[1.69\times 10^{19}]{g}$.

\begin{figure}
\centering
\includegraphics[scale=0.55]{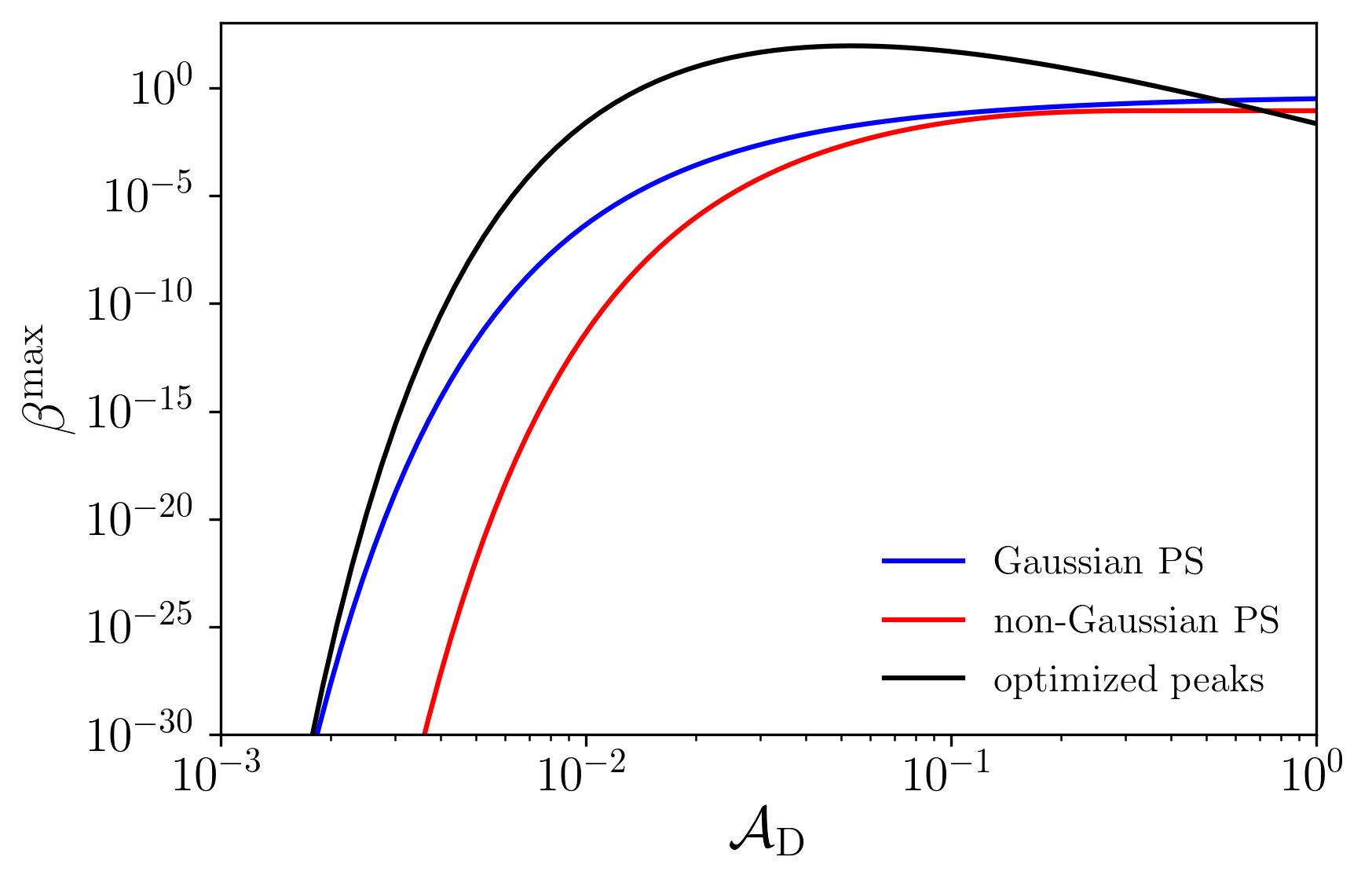}
\caption{Variation of peak value of abundance $\beta^{\text{max}}$ with amplitude}
\end{figure}

\subsection{Extended $\mathcal{P}_{\zeta}$}
To compare the Press-Schecter results with peaks, the extended power spectrum in \cite{Yoo:2018kvb} is used. Although it is not a lognormal function like the one used in Sec. (\ref{sec:lognormal}), one can verify that the function
\begin{equation}\label{eq:extended_power}
\mathcal{P}_{\zeta}(k)=3\sqrt{\frac{6}{\pi}}\sigma_{0}^{2}\left( \frac{k}{k_{\star}} \right)^{3}\exp\left( -\frac{3}{2}\frac{k^{2}}{k_{\star}^{2}} \right)
\end{equation}
approximates a lognormal distribution of $\sigma=0.5$ rather well. Another benefit of using a Gaussian function of this form is that the $n$-th moments $\sigma_{n}^{2}$ and $\psi(r)$ can be calculated analytically.\footnote{One shortcoming of using such a Gaussian function is that the ease of modulating the width is lost. The width can be controlled by using higher powers of the $\frac{k}{k_{\star}}$ prefactor. However, this subsequently creates very complicated equations that need to be dealt with.} For the power spectrum given by Eq. \eqref{eq:extended_power}, we have $k_{c}=k_{\star},\gamma=\sqrt{3/5}$ and
\begin{align*}
\sigma_{n}^{2}&=\frac{2^{n+1}}{3^{n}\sqrt{\pi}}\Gamma\left( \frac{3}{2}+n \right)\sigma_{0}^{2}k_{\star}^{2n}\\
g(r;k_{*})&=-\frac{1}{6}\exp\left( -\frac{1}{6}k_{\star}^{2}r^{2} \right)\left[ 6+k_{\star}^{2}r^{2}\left( 1-\frac{k_{*}^{2}}{k_{\star}^{2}} \right) \right] \numberthis
\end{align*}

\begin{figure}
\centering
\includegraphics[scale=0.55]{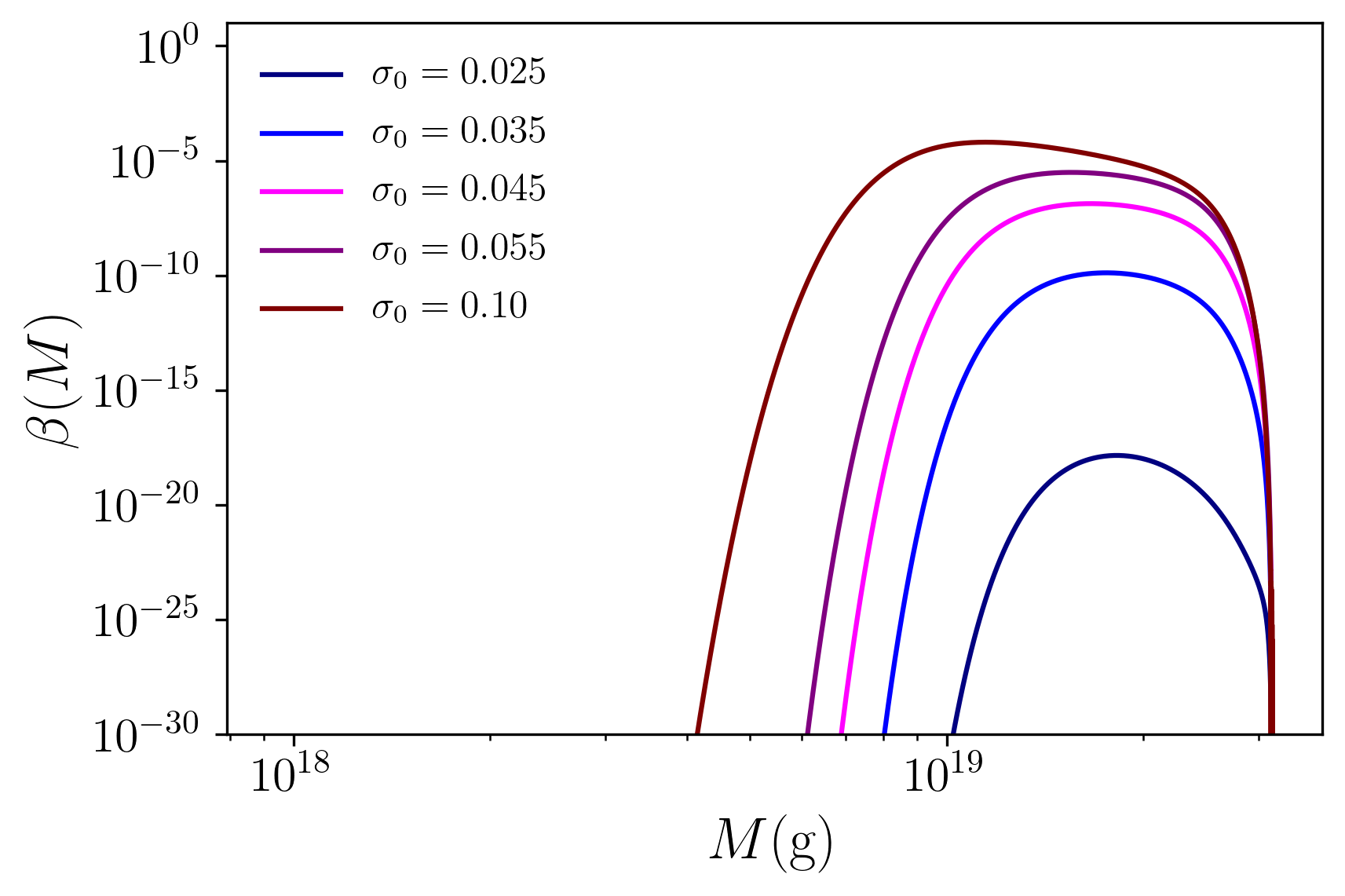}
\includegraphics[scale=0.55]{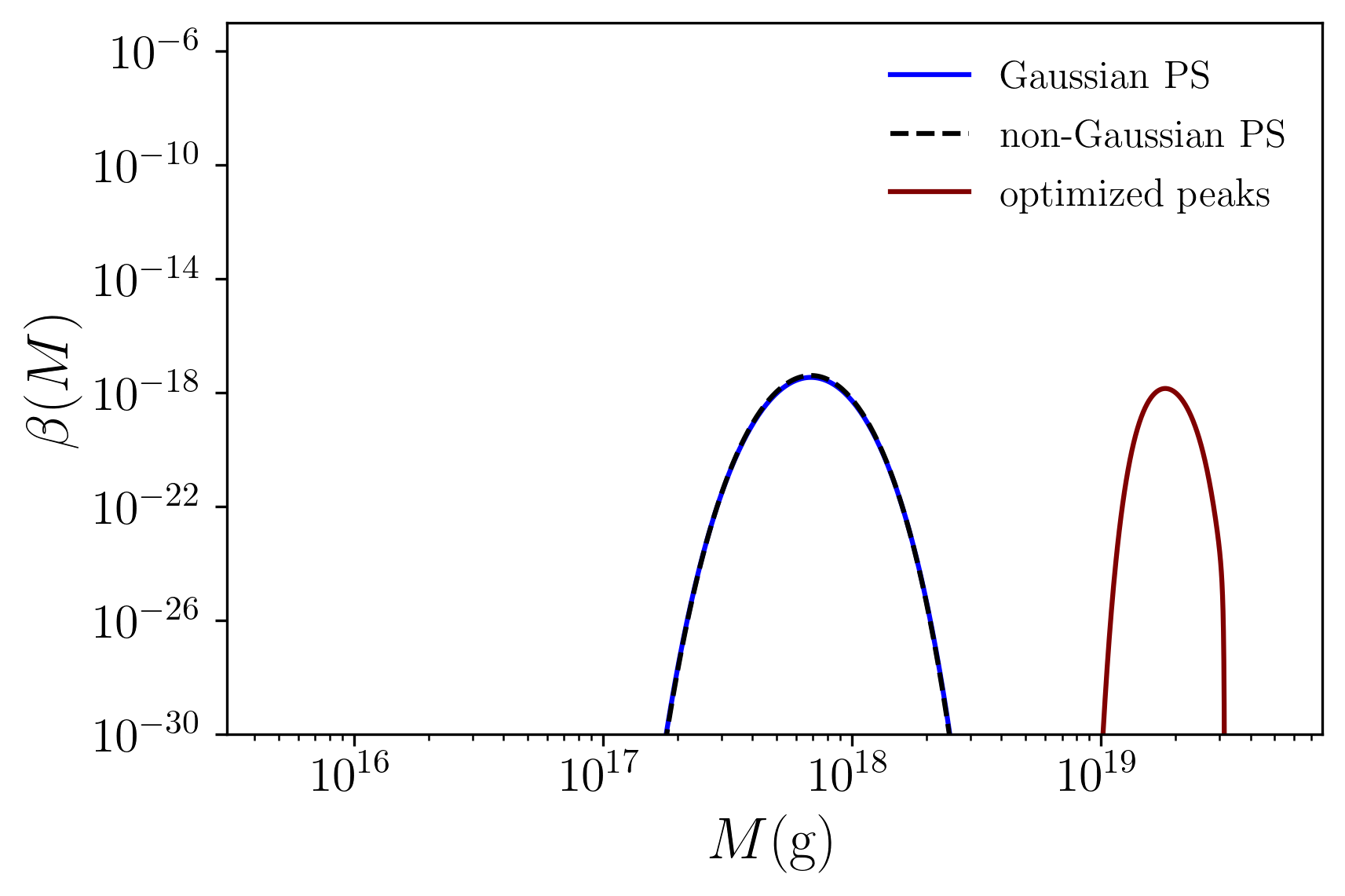}
\caption{\textit{Left panel}: PBH abundance using peaks theory for extended $\mathcal{P}_{\zeta}$ and different values of $\sigma_{0}$; \textit{Right panel}: Comparison with Press-Schechter}
\label{fig:peaks_extended}
\end{figure}
An exact expession for the formation fraction cannot be obtained and $\beta$ must be computed using Eq. \eqref{eq:beta_peaks}. The procedure is described in detail in the original paper and interested readers are encouraged to go through it. Skipping these details, the results are shown in Fig. (\ref{fig:peaks_extended}). The figure on the left shows the dependence of $\beta(M)$ on $\sigma_{0}$ (and, hence, the amplitude). The figure on the right shows a comparison of the peaks calculation with the modified Press-Schechter. Here, for the lognormal Press-Schechter calculation $\mathcal{A}_{\text{LN}}=0.006$ and $\sigma=0.5$. To maintain the same PBH abundance as the Press-Schechter case, the amplitude of the extended $\mathcal{P}_{\zeta}$ is set such that $\sigma_{0}=0.025$ while the same threshold overdensity $\bar{\delta}_{\text{th}}\sim 0.55$ has been used.\\\\
 The optimized peaks calculation allows PBHs to be formed at the required abundance with power spectra having smaller amplitudes. The shift towards larger mass also has consequences for inflationary model building concerned with PBH formation. For example, in \cite{dalianis,Mahbub:2019uhl}, the inflationary $\alpha-$attractor model was used. The inflaton potentials that were considered in these papers produced PBHs in the mass range $10^{17}-\unit[10^{18}]{g}$. However, the predicted values of $n_{s}$ at the CMB scale were somewhat smaller than expected. One way to circumvent this problem (atleast from a modelling point of view) would be to consider inflation that lasts for more $e$-folds. This is not without problems, as an increase in $e$-folds would inadvertantly push the peak in $\mathcal{P}_{\zeta}$ to higher values of $k$ and, hence, towards lower mass. However, the mass calculation using Eq. \eqref{eq:new_PBH_mass} can aid in a re examination  of such inflationary models and help improve the value of $n_{s}$ while also producing PBHs that are cosmologically relevant in the current epoch.

\section{Conclusion}
In this paper, the impact of the nonlinear statistics of overdensities on PBH abundance has been explored. The analytical methods that were developed in \cite{Kawasaki:2019mbl} to study the LIGO-PBH scenario using Dirac delta function $\mathcal{P}_{\zeta}$ were adapted to study PBH abundance in the $10^{17}-10^{18}\text{g}$ mass range using $\mathcal{P}_{\zeta}$ modelled as both Dirac delta and lognormal functions. Moreover, these were used to study an inflationary $\alpha-$attractor model that was previously worked upon. Also, the nonvanishing kurtosis was derived, which turned out to be positive. The results were consistent for both Dirac delta and lognormal power spectra in that, for the negative skewness case, the non-Gaussian abundance becomes of the order of the Gaussian one when $\mathcal{A}_{\text{NG}}\simeq1.4^{2}\mathcal{A}_{\text{G}}$ while, for the positive kurtosis case, it turned out to be $\mathcal{A}_{\text{NG}}\simeq0.92\mathcal{A}_{\text{G}}$. \\\\
\indent The combined effect of $\tilde{\kappa}_{3}<0$ and $\tilde{\kappa}_{4}>0$ on the abundance has been briefly explored in Appendix (\ref{appendix:B}), which introduces smaller suppression in the PBH abundance as one would expect. Of course, higher order cumulants can be derived in a similar fashion but, seeing that these will likely depend on higher powers of $\sigma$, might not induce significant effects on $\beta$. Indeed, it can be seen that the effect of $\tilde{\kappa}_{4}$ was not as severe and might lend credence to the fact that $\sigma$ might act as a small parameter in the expansion in Eq. \eqref{eq:delprime}. \\\\
\indent Some other differences also need to be elucidated. As mentioned in Sec. (\ref{sec:overdensity_statistics}), the spherical decomposition automatically introduces a top hat window function whereas, in the study of PBH formation, a Gaussian window function had been mostly employed. As a result, calculations using the latter usually reported that a higher $\mathcal{P}_{\zeta}$ is required for producing PBHs at an appreciable abundance. The numbers reported were usually $\mathcal{P}_{\zeta}\sim\mathcal{O}(10^{-1})$ for $\delta_{\text{th}}\sim0.4-0.5$. Another likely source of uncertainties arises from the fact that the correspondance $k\sim r_{m}^{-1}$ was made in the PBH mass calculations which is not entirely accurate. The mass is more directly related to $R(r_{m})$ and such a simplification should introduce deviations from a fully numerical computation \cite{Kawasaki:2019mbl,Tada:2019amh}. This fact is reflected when the Press-Schechter results have been compared to the optimized peaks calculation, where it has been shown that $\beta(M)$ is centered around masses which are typically an order of magnitude larger than simplified calculations with Eq. \eqref{eq:PBH_mass}. With the optimized peaks approach, PBH formation is more efficient and one does not not need to worry about the choice of window function.

\section*{Acknowledgement}
The author is grateful to A. De and J. Kapusta for their useful comments on the manuscript. The author also thanks the anonymous referee who suggested the inclusion of the peaks theory analysis and other helpful comments and C.-M. Yoo for email correspondance regarding peaks theory calculations.

\appendix
\section{Calculation of skewness and kurtosis of the overdensities}
\label{appendix:A}
Here the expressions for skewness and kurtosis are derived. We start with the expressions for $\bar{\delta}^{(1)}$ and $\bar{\delta}^{(2)}$ from Eq. \eqref{eq:del1} and \eqref{eq:del2}. The terms in the skewness are expanded out in terms of $\bar{\delta}^{(1,2)}$ with the only nonvanishing contributions being\footnote{The only nonvanishing terms are the ones which have an even number of $\zeta$ inside the expectation values. Thus, $\langle \zeta_{1}\zeta_{2}\cdot\cdot\cdot\zeta_{2n+1} \rangle=0$. As a result, in the calculations, $\bar{\delta}^{(1)}$ appears with a square to produce its lowest order contribution since it contains only one factor of $\zeta$.}
\begin{equation}
\tilde{\kappa}_{3}(r_{m})\simeq \frac{3}{\sigma(r_{m})^{3}}\left( \langle \bar{\delta}^{(1)}(r_{m})^{2}\bar{\delta}^{(2)}(r_{m}) \rangle -\langle \bar{\delta}^{(1)}(r_{m})^{2} \rangle \langle \bar{\delta}^{(2)}(r_{m}) \rangle \right) +\mathcal{O}(\zeta^{6})
\end{equation}
Then, 
\begin{multline}\label{eq:one}
\langle \bar{\delta}^{(1)}(r_{m})^{2}\bar{\delta}^{(2)}(r_{m}) \rangle=-\left( \frac{4}{9} \right)^{3}\frac{1}{6}\frac{1}{(2\pi^2)^2}\int_{0}^{\infty}\frac{dk_{1}}{k_{1}}\frac{dk_{2}}{k_{2}}\frac{dk_{3}}{k_{3}}\frac{dk_{4}}{k_{4}}(r_{m}^{_-2})^{4}(k_{1}r_{m})^{4}(k_{2}r_{m})^{4}(k_{3}r_{m})^{4}(k_{4}r_{m})^{4} \\
\times W(k_{1}r_{m})W(k_{2}r_{m})W(k_{3}r_{m})W(k_{4}r_{m})\langle \zeta(k_{1})\zeta(k_{2})\zeta(k_{3})\zeta(k_{4}) \rangle
\end{multline}

\begin{multline}\label{eq:two}
\langle \bar{\delta}^{(1)}(r_{m})^{2}\rangle \langle\bar{\delta}^{(2)}(r_{m}) \rangle=-\left( \frac{4}{9} \right)^{3}\frac{1}{6}\frac{1}{(2\pi^2)^2}\int_{0}^{\infty}\frac{dk_{1}}{k_{1}}\frac{dk_{2}}{k_{2}}\frac{dk_{3}}{k_{3}}\frac{dk_{4}}{k_{4}}(r_{m}^{_-2})^{4}(k_{1}r_{m})^{4}(k_{2}r_{m})^{4}(k_{3}r_{m})^{4}(k_{4}r_{m})^{4} \\
\times W(k_{1}r_{m})W(k_{2}r_{m})W(k_{3}r_{m})W(k_{4}r_{m})\langle \zeta(k_{1})\zeta(k_{2})\rangle  \langle\zeta(k_{3})\zeta(k_{4}) \rangle
\end{multline}
It can be seen that both Eq. \eqref{eq:one} and \eqref{eq:two} are the same except for the $\zeta$ correlation functions. Now, a $2n-$point correlation function of the type $\langle \zeta_{k_{1}}\zeta_{k_{2}}\cdot\cdot\cdot\zeta_{k_{2n}} \rangle$ can be expressed in terms of products of two point correlation functions \cite{Bartolo:2004if} (physicists call it Wick's theorem while statisticians/probabilists call it Isserlis' theorem). There will also be a connected term that usually encodes information related to the bispectrum, trispectrum and higher order contributions\footnote{The connected correlation functions are relevant if primordial non-Gaussianity is under consideration.} which will be ignored here. The four point correlation function can be decomposed into 
\begin{align*}
\langle \zeta(k_{1})\zeta(k_{2})\zeta(k_{3})\zeta(k_{4}) \rangle&=\langle \zeta(k_{1})\zeta(k_{2}) \rangle \langle \zeta(k_{3})\zeta(k_{4}) \rangle + \langle \zeta(k_{1})\zeta(k_{3}) \rangle \langle \zeta(k_{2})\zeta(k_{4}) \rangle + \langle \zeta(k_{1})\zeta(k_{4}) \rangle \langle \zeta(k_{2})\zeta(k_{3}) \rangle\\
&= \frac{2\pi^2}{k_{1}^3}\delta(k_{1}-k_{2})\mathcal{P}_{\zeta}(k_{1}) +\text{2 perms} \numberthis
\end{align*}
From here, it is a matter of counting and it can be seen that two of these integrals contribute to the skewness. Then
\begin{align*}
\tilde{\kappa}_{3}(r_{m})\sigma(r_{m})^{3}&=-\left( \frac{4}{9} \right)^{3}\int_{0}^{\infty}\frac{dk_{1}}{k_{1}}\frac{dk_{2}}{k_{2}}(k_{1}r_{m})^{4}(k_{2}r_{m})^{4}W(k_{1}r_{m})^{2}W(k_{2}r_{m})^{2}\mathcal{P}_{\zeta}(k_{1})\mathcal{P}_{\zeta}(k_{2})\\
&=-\frac{9}{4}\sigma(r_{m})^{4}\\
\tilde{\kappa}_{3}(r_{m})&=-\frac{9}{4}\sigma(r_{m}) \numberthis
\end{align*}
The kurtosis can be similarly calculated. Expanding out the terms
\begin{multline}
\tilde{\kappa}_{4}(r_{m})\simeq\frac{1}{\sigma(r_{m})^{4}}\biggl( \langle \bar{\delta}^{(1)}(r_{m})^{4} \rangle + 6\langle \bar{\delta}^{(1)}(r_{m})^{2}\bar{\delta}^{(2)}(r_{m})^{2} \rangle -12\langle \bar{\delta}^{(1)}(r_{m})^{2}\bar{\delta}^{(2)}(r_{m}) \rangle \langle \bar{\delta}^{(2)}(r_{m}) \rangle \\
+12 \langle \bar{\delta}^{(1)}(r_{m})^{2} \rangle \langle \bar{\delta}^{(2)}(r_{m}) \rangle^{2} -3\langle \bar{\delta}^{(1)}(r_{m})^{2} \rangle^{2} +6\langle \bar{\delta}^{(1)}(r_{m})^{2} \rangle\langle \bar{\delta}^{(2)}(r_{m})^{2} \rangle \biggr)+\mathcal{O}(\zeta^{8})
\end{multline}
where $\tilde{\kappa}_{4}$ now contains terms that are proportional to $\langle \zeta\zeta\zeta\zeta \rangle$, $\langle\zeta\zeta\zeta\zeta\zeta\zeta\rangle$, $\langle \zeta\zeta\zeta\zeta \rangle \langle \zeta\zeta \rangle$ and $\langle \zeta\zeta \rangle\langle \zeta\zeta \rangle\langle \zeta\zeta \rangle$. Since the first term containing the four point correlation function will produce three permutations, they will cancel out the fifth term inside the parentheses.\footnote{It is true that, apart from the $\zeta$ correlation functions, the integrals are similar. After expanding out the correlation functions, the terms in $\tilde{\kappa}_{4}$ are all integrated over three momentum measures containing three factors of the window function and power spectrum $$ \tilde{\kappa}_{4}\sim \int\prod_{i=1}^{3}\frac{dk_{i}}{k_{i}}(k_{i}r_{m})^{2}W(k_{i}r_{m})^{2}\mathcal{P}_{\zeta}(k_{i})$$ Other higher order cumulants are generalizations of this where the coefficients will be determined by the number of such integrals appearing in $\tilde{\kappa}_{n}$.} Now, schematically, the remaining $\mathcal{O}(\zeta^{6})$ terms will look like
\begin{align*}
\langle \bar{\delta}^{(1)}(r_{m})^{2}\bar{\delta}^{(2)}(r_{m})^{2} \rangle &\sim\langle \zeta(k_{1})\zeta(k_{2})\zeta(k_{3})\zeta(k_{4})\zeta(k_{5})\zeta(k_{6}) \rangle\\
&\sim (2\pi^2)^{3}\left[ \frac{1}{k_{1}^{3}}\delta(k_{1}-k_{2})\mathcal{P}_{\zeta}(k_{1})\frac{1}{k_{3}^{3}}\delta(k_{3}-k_{4})\mathcal{P}_{\zeta}(k_{3})\frac{1}{k_{5}^{3}}\delta(k_{5}-k_{6})\mathcal{P}_{\zeta}(k_{5}) +\text{14 perms}\right]\numberthis
\end{align*}

\begin{align*}
\langle \bar{\delta}^{(1)}(r_{m})^{2}\bar{\delta}^{(2)}(r_{m}) \rangle \langle \bar{\delta}^{(2)}(r_{m}) \rangle &\sim\langle \zeta(k_{1})\zeta(k_{2})\zeta(k_{3})\zeta(k_{4}) \rangle \langle \zeta(k_{5})\zeta(k_{6}) \rangle \\
&\sim (2\pi^2)^{3}\biggl[ \frac{1}{k_{1}^{3}}\delta(k_{1}-k_{2})\mathcal{P}_{\zeta}(k_{1})\frac{1}{k_{3}^{3}}\delta(k_{3}-k_{4})\mathcal{P}_{\zeta}(k_{3})\frac{1}{k_{5}^{3}}\delta(k_{5}-k_{6})\mathcal{P}_{\zeta}(k_{5}) \\
& \qquad \;\;\;\;\;\;\;+\text{2 perms}\biggr]\numberthis
\end{align*}

\begin{align*}
\langle \bar{\delta}^{(1)}(r_{m})^{2}\rangle \langle \bar{\delta}^{(2)}(r_{m}) \rangle^{2}&\sim \langle \zeta(k_{1})\zeta(k_{2}) \rangle \langle \zeta(k_{3})\zeta(k_{4}) \rangle \langle \zeta(k_{5})\zeta(k_{6}) \rangle \\
&\sim (2\pi^2)^{3}\left[ \frac{1}{k_{1}^{3}}\delta(k_{1}-k_{2})\mathcal{P}_{\zeta}(k_{1})\frac{1}{k_{3}^{3}}\delta(k_{3}-k_{4})\mathcal{P}_{\zeta}(k_{3})\frac{1}{k_{5}^{3}}\delta(k_{5}-k_{6})\mathcal{P}_{\zeta}(k_{5})\right]
\end{align*}
Again, it becomes a matter of counting how many of these integrals are present in the computation and one can see that there will be 48 of them. Expressing the kurtosis in a manner similar to that of the skewness
\begin{align*}
\tilde{\kappa}_{4}(r_{m})\sigma(r_{m})^{4}&=3\cdot\left( \frac{4}{9} \right)^{5}\int_{0}^{\infty}\frac{dk_{1}}{k_{1}}\frac{dk_{2}}{k_{2}}\frac{dk_{3}}{k_{3}}(k_{1}r_{m})^{4}(k_{2}r_{m})^{4}(k_{3}r_{m})^{4} W(k_{1}r_{m})^{2}W(k_{2}r_{m})^{2}W(k_{3}r_{m})^{2}\\
&\qquad \;\;\;\;\;\;\;\;\;\;\;\;\;\;\;\;\;\;\;\times\mathcal{P}_{\zeta}(k_{1})\mathcal{P}_{\zeta}(k_{2})\mathcal{P}_{\zeta}(k_{3})\\
&=3\cdot\frac{9}{4}\sigma(r_{m})^{6}\\
\tilde{\kappa}_{4}(r_{m})&=\frac{27}{4}\sigma(r_{m})^{2}
\end{align*}
Hence, much like the skewness, the kurtosis simplifies into a function of the variance. However, the kurtosis is positive which means that the area under the tail of the PDF is greater which would imply greater PBH production. Much like the same way, even higher order cumulants may be calculated although, if they are higher powers of $\sigma^{2}$, their contributions will likely be suppressed.

\section{Combined effect of $\tilde{\kappa}_{3}<0$ and $\tilde{\kappa}_{4}>0$}
\label{appendix:B}
\begin{figure}[h]
\includegraphics[scale=0.55]{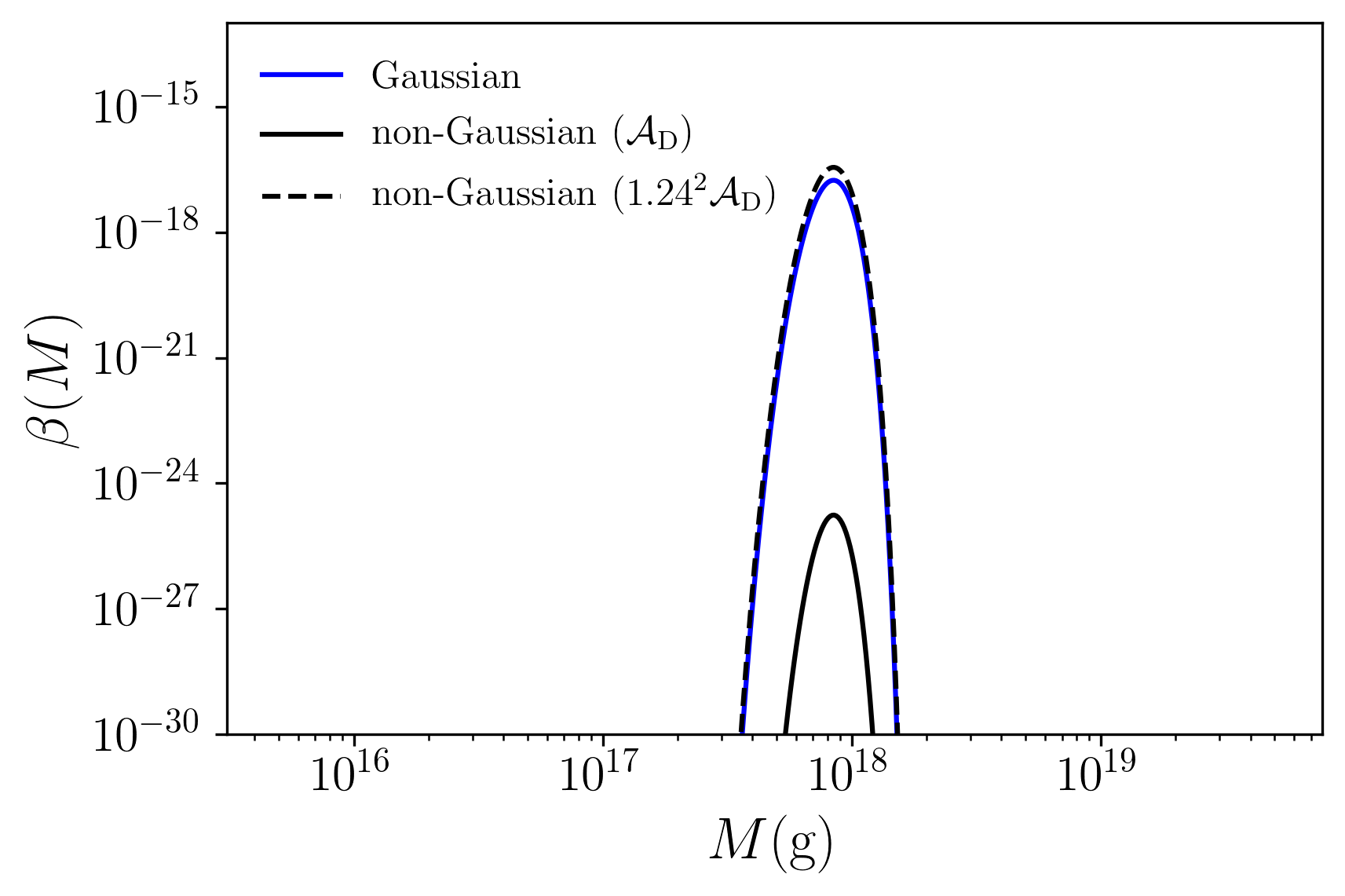}
\includegraphics[scale=0.55]{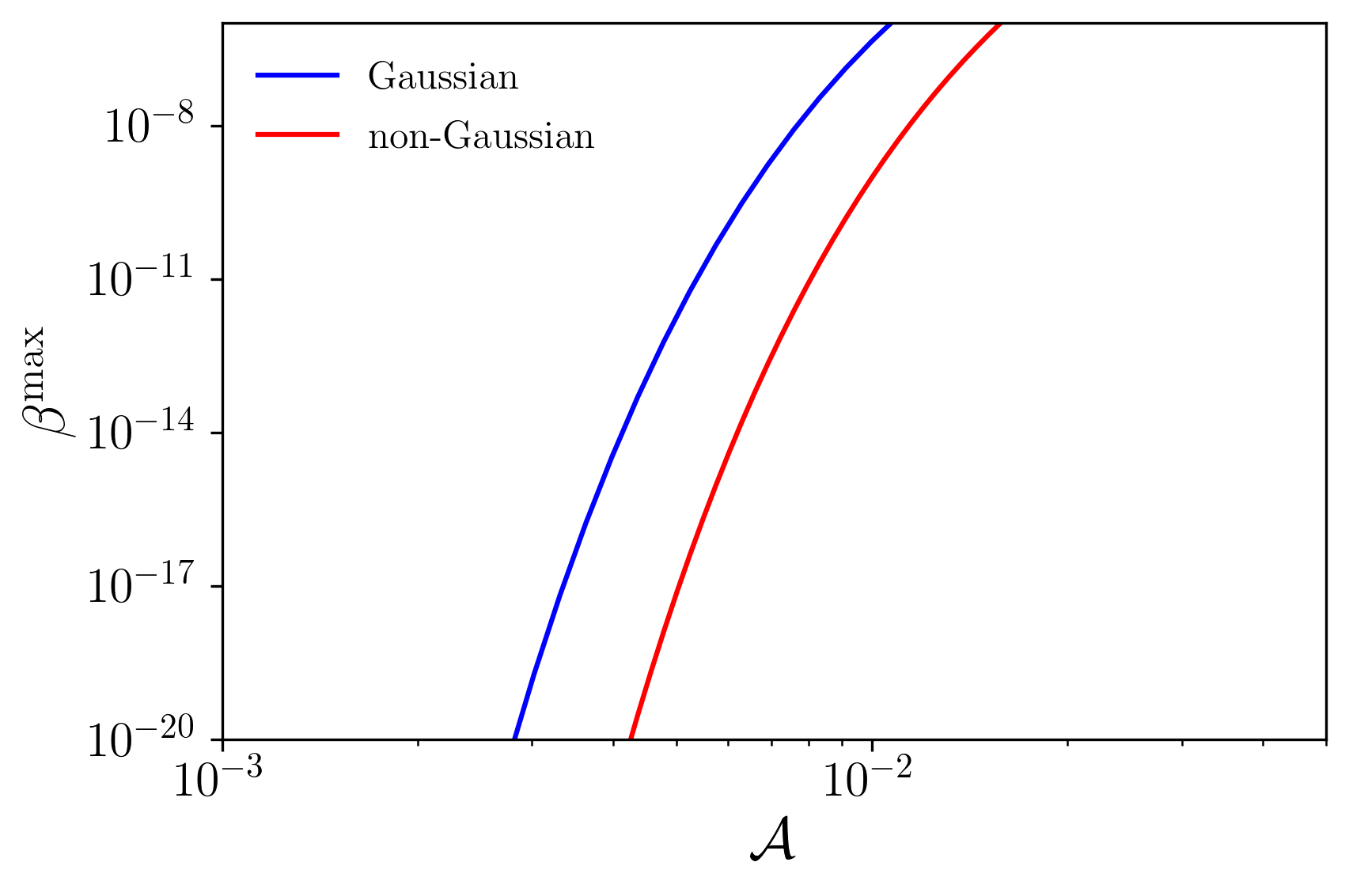}
\caption{\textit{Left panel}: PBH abundance $\beta(M)$ at formation for Gaussian overdensities and non-Gaussian overdensities with two different amplitudes for Dirac delta $\mathcal{P}_{\zeta}$ with $\tilde{\kappa}_{3}<0$ and $\tilde{\kappa}_{4}>0$; \textit{Right panel}: Variation of the peak value of abundance $\beta^{\text{max}}$ with amplitude}
\label{fig:dirac_delta_stuff_blah}
\end{figure}

It would be of interest to observe the combined effect of the nonvanishing skewness and kurtosis on the PBH abundance. Here, it is done for the Dirac delta function power spectrum given by Eq. \eqref{eq:delta_power}. As in Sec. (\ref{sec:PBH_abundance}), the starting point would be to invert Eq. \eqref{eq:delprime} for $\tilde{\kappa}_{3},\tilde{\kappa}_{4}\neq0$. The resulting cubic equation has only one real solution
\begin{multline*}
\delta_{\text{G}}=-\frac{18\tilde{\kappa}_{3}\sigma^{3}}{\tilde{\kappa}_{4}^{2}}+\frac{2^{1/3}}{3\tilde{\kappa}_{4}^{2}}X(\sigma,\tilde{\kappa}_{3},\tilde{\kappa}_{4})\left[ Y(\sigma,\tilde{\kappa}_{3},\tilde{\kappa}_{4})+\sqrt{4X(\sigma,\tilde{\kappa}_{3},\tilde{\kappa}_{4})^{3}+Y(\sigma,\tilde{\kappa}_{3},\tilde{\kappa}_{4})^{2}} \right]^{-1/3} \\
 -\frac{1}{2^{1/3}\cdot 3\tilde{\kappa}_{4}^{2}}\left[ Y(\sigma,\tilde{\kappa}_{3},\tilde{\kappa}_{4})+\sqrt{4X(\sigma,\tilde{\kappa}_{3},\tilde{\kappa}_{4})^{3}+Y(\sigma,\tilde{\kappa}_{3},\tilde{\kappa}_{4})^{2}} \right]^{1/3} \numberthis
\end{multline*}
where
\begin{align}
X(\sigma,\tilde{\kappa}_{3},\tilde{\kappa}_{4})&=972\sigma^{4}\tilde{\kappa}_{4}^{2}-2916\sigma^{6}\tilde{\kappa}_{3}^{2} \\
Y(\sigma,\tilde{\kappa}_{3},\tilde{\kappa}_{4})&=-8748\bar{\delta}\sigma^{4}\tilde{\kappa}_{4}^{4}-1458\sigma^{5}\tilde{\kappa}_{3}\tilde{\kappa}_{4}^{4}-157464\sigma^{7}\tilde{\kappa}_{3}\tilde{\kappa}_{4}^{2}+314928\sigma^{9}\tilde{\kappa}_{3}^{3}
\end{align}
The results for the Dirac delta function power spectrum are given in Fig. (\ref{fig:dirac_delta_stuff_blah}) using the same values of $k_{\star}$ and $\mathcal{A}_{\text{D}}$. The results conclude that when the skewness and kurtosis are combined, the effect is intermediate compared to the case where skewness and kurtosis are considered separately. The non-Gaussian abundance becomes comparable to the Gaussian one when the amplitude is increased to $1.24^{2}\mathcal{A}_{\text{D}}$. The calculations can be repeated for the lognormal power spectrum as well and one can find out that a similar adjustment in the amplitude will result in comparable abundances. Likewise, in the $\beta^{\text{max}}$ plot, the non-Gaussian curve will tend to shift towards decreasing $\mathcal{A}_{\text{LN}}$.

\section{Effects of choice of window function on smoothing}
\label{appendix:C}

It is common practice in cosmology and structure formation to consider the overdensity perturbations smoothed to a certain length scale before the Press-Schechter theory is applied to assign masses to such gravitationally bound objects. This smoothing procedure essentially tries to remove perturbations at length scales smaller than that is relevant. Hence, there is often an ambiguity on what type of smooting (filter) function to use. Widely used in literature are the (i) real space top hat and (ii) Gaussian filter functions. In Fourier space, these filter functions take the form \cite{Liddle:2000cg,1993sfu..book.....P}

\[
  W(z) =
  \begin{cases}
                                   3\left( \frac{\sin z}{z^3}-\frac{\cos z}{z^2} \right) & \text{Top Hat} \\
  \exp\left( -\frac{z^2}{2} \right) & \text{Gaussian}\numberthis
  \end{cases}
\]
where $z=kR$ with $R$ having dimensions of length. These two filter functions behave rather differently when it comes to dealing with subhorizon modes. To illustrate this, let us consider a nearly scale-invariant curvature power spectrum, typically favoured by slow-roll inflation, of the form

\begin{equation}
\mathcal{P}_{\zeta}(k)=\mathcal{A}_{s}\left( \frac{k}{0.05} \right)^{n_{s}-1}
\end{equation}
where $\mathcal{A}_{s}=2.2\times 10^{-9}$ is the CMB normalization and $n_{s}\approx 0.965$ is the scalar spectral index \cite{Akrami:2018odb}. Now considering the power spectrum of overdensity perturbations smoothed by a filter function
\begin{equation}\label{eq:powe_overdensity}
\mathcal{P}_{\delta}\left(\frac{k}{aH}\right)=\frac{16}{81}\left( \frac{k}{aH} \right)^{4}W\left(\frac{k}{aH}\right)^{2}\mathcal{P}_{\zeta}(k)
\end{equation}
Eq. \eqref{eq:powe_overdensity} formally appears in the definition of the variance of the distribution of overdensities.

\begin{figure}[h]
\centering
\includegraphics[scale=0.5]{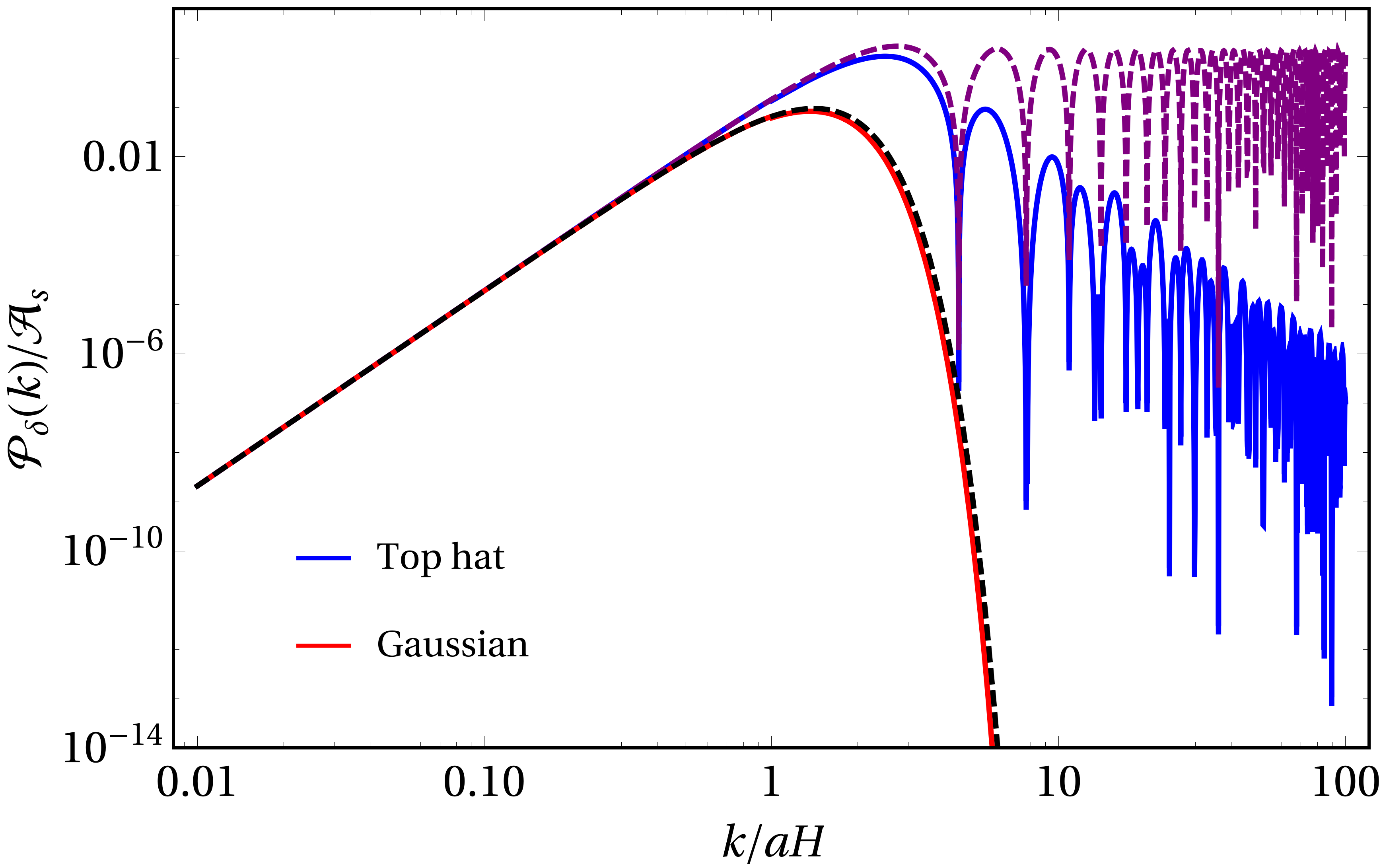}
\caption{Effect of filter functions on the power spectrum of overdensities $\mathcal{P}_{\delta}$ for the nearly scale-invariant curvature power spectrum. The dashed lines represent the effects of not including a linear transfer function.}
\label{fig:window}
\end{figure}

The effects of the two filter functions are shown in Fig. (\ref{fig:window}). The plot legends describe the power spectrum smoothed by the top hat and Gaussian filters with the inclusion of the linear transfer function. Fixing the horizon scale at $aH=1$, it can be seen that, while the Gaussian filter nicely removes subhorizon contributions, the top hat function is not so efficient. In fact, without the transfer function, the effect of the top hat function is even more severe (shown in dotted magenta line). One immediate consequence of this can be found in the variance of the overdensities defined as
\begin{equation}
\sigma^{2}(q)=\int d\ln k\;\mathcal{P}_{\delta}(k,q^{-1})
\end{equation}
Because of the subhorizon contributions $\sigma^{2}_{\text{TH}}(q)>>\sigma^{2}_{\text{G}}(q)$. This, as a result, quite adversely affects the PBH formation fraction, which is dependent on $\sigma^{2}$.


\bibliography{pbh_nonlin}

\end{document}